\documentclass[amsmath,amssymb,aps,10pt,prd,twocolumn,letterpaper,nofootinbib,balancelastpage,notitlepage,superscriptaddress,floatfix,preprintnumbers]{revtex4-2}
\usepackage{graphicx}	
\usepackage{ragged2e}	
\usepackage{bm}		    
\usepackage{soul}
\usepackage[sort&compress]{natbib}	 	
	\setcitestyle{square,numbers,comma}	
\usepackage[colorlinks=true,urlcolor=blue,linkcolor=red,citecolor=red]{hyperref}	
\newcommand{\tabref}[2][]{Tab{#1}.~\ref{tab:#2}}		
\newcommand{\figref}[2][]{Fig{#1}.~\ref{fig:#2}}		
\newcommand{\secref}[2][]{Sec{#1}.~\ref{sec:#2}}		
\newcommand{\appref}[2][x]{Appendi{#1}~\ref{app:#2}}	
\renewcommand{\eqref}[2][]{Eq{#1}.~(\ref{eq:#2})}		
\newcommand{\citeR}[2][]{Ref{#1}.~\cite{#2}}			
\newcommand{\dd}[2]{\frac{\partial #1}{\partial #2}}
\newcommand{\ddd}[2]{\frac{\partial^2#1}{\partial #2^2}}
\newcommand{\DM}{\mathrm{DM}}
\newcommand{\LL}{\mathcal L}
\newcommand{\N}{\bm{\hat n}}
\newcommand{\nl}{\nonumber \\ & \quad }					
\newcommand{\PHI}{\bm{\hat\phi}}
\newcommand{\THETA}{\bm{\hat\theta}}
\newcommand{\Z}{\bm{\hat z}}

\begin{document}

\title{Ultralight dark matter detection with levitated ferromagnets}
\date{\today}
\author{Saarik Kalia}
\email{kalias@umn.edu}
\affiliation{School of Physics \& Astronomy, University of Minnesota, Minneapolis, MN 55455, USA}
\author{Dmitry Budker}
\affiliation{Johannes Gutenberg University Mainz, 55128 Mainz, Germany}
\affiliation{Helmholtz-Institute, GSI Helmholtzzentrum fur Schwerionenforschung, 55128 Mainz, Germany}
\affiliation{Department of Physics, University of California at Berkeley, Berkeley, California 94720-7300, USA}
\author{Derek F.~Jackson Kimball}
\affiliation{Department of Physics, California State University -- East Bay, Hayward, CA 94542, USA}
\author{Wei Ji}
\affiliation{Johannes Gutenberg University Mainz, 55128 Mainz, Germany}
\affiliation{Helmholtz-Institute, GSI Helmholtzzentrum fur Schwerionenforschung, 55128 Mainz, Germany}
\author{Zhen Liu}
\affiliation{School of Physics \& Astronomy, University of Minnesota, Minneapolis, MN 55455, USA}
\author{Alexander O.~Sushkov}
\affiliation{Department of Physics, Boston University, Boston, MA 02215, USA}
\affiliation{Department of Electrical and Computer Engineering, Boston University, Boston, MA 02215, USA}
\affiliation{Photonics Center, Boston University, Boston, MA 02215, USA}
\author{Chris Timberlake}
\affiliation{School of Physics and Astronomy, University of Southampton, Southampton SO17 1BJ, UK}
\author{Hendrik Ulbricht}
\affiliation{School of Physics and Astronomy, University of Southampton, Southampton SO17 1BJ, UK}
\author{Andrea Vinante}
\affiliation{Istituto di Fotonica e Nanotecnologie IFN--CNR, 38123 Povo, Trento, Italy}
\affiliation{Fondazione Bruno Kessler (FBK), 38123 Povo, Trento, Italy}
\author{Tao Wang}
\affiliation{A*STAR Quantum Innovation Centre (Q.InC), Institute of Materials Research and Engineering (IMRE), Agency for Science, Technology and Research (A*STAR), 2 Fusionopolis Way, 08-03, Singapore, 138634, Republic of Singapore}

\preprint{UMN-TH-4329/24}
\preprint{FERMILAB-PUB-24-0545-SQMS}

\begin{abstract}
Levitated ferromagnets act as ultraprecise magnetometers, which can exhibit high quality factors due to their excellent isolation from the environment.  These instruments can be utilized in searches for ultralight dark matter candidates, such as axionlike dark matter or dark-photon dark matter. In addition to being sensitive to an axion-photon coupling or kinetic mixing, which produce physical magnetic fields, ferromagnets are also sensitive to the effective magnetic field (or ``axion wind") produced by an axion-electron coupling.  While the dynamics of a levitated ferromagnet in response to a DC magnetic field have been well studied, all of these couplings would produce AC fields.  In this work, we study the response of a ferromagnet to an applied AC magnetic field and use these results to project their sensitivity to axion and dark-photon dark matter.  We pay special attention to the direction of motion induced by an applied AC field, in particular, whether it precesses around the applied field (similar to an electron spin) or librates in the plane of the field (similar to a compass needle).  We show that existing levitated ferromagnet setups can already have comparable sensitivity to an axion-electron coupling as comagnetometer or torsion balance experiments.  In addition, future setups can become sensitive probes of axion-electron coupling, dark-photon kinetic mixing, and axion-photon coupling, for ultralight dark matter masses $m_\DM\lesssim\mathrm{feV}$.
\end{abstract}

\maketitle

\section{Introduction}
\label{sec:introduction}

Levitated ferromagnets can serve as excellent instruments for precision measurements of torques and magnetic fields~\cite{jacksonkimball2016,prat-camps2017,vinante2020,vinante2021,ahrens2024}, which can be applied to tests of fundamental physics and searches for new physics~\cite{fadeev2021,gps2021}.  Due to the intrinsic spin of its polarized electrons, a ferromagnet may act as a gyroscope in the limit where the spin contribution $S=N\hbar/2$ to its total angular momentum dominates over the contribution from its rotational angular momentum $L=I\omega$~\cite{jacksonkimball2016}.%
\footnote{Generically, this is a tensor relation.  For simplicity, here we assume the moment of inertia tensor $I$ is diagonal; see also discussion following \eqref{rotation}.}
In such a case, the ferromagnet will precess around an applied DC magnetic field, similar to a single electron spin.  In the opposite limit $S\ll L$, the dominant motion of the ferromagnet will be to librate in the plane of the applied field, similar to a compass needle.

In order to realize the potential of this system, the ferromagnet must be adequately isolated from its environment.  One of the most promising ways is to levitate the ferromagnet over a superconducting plane~\cite{wang2019,vinante2020,fadeev2021,vinante2021,ahrens2024}.  In such a scenario, the ferromagnet is repelled by an ``image" magnetic dipole located below the plane.  The presence of this superconducting plane can significantly affect the dynamics of the ferromagnet, slowing down its precession frequency.%
\footnote{As discussed in \citeR{fadeev2021}, a ferromagnet levitated above a superconducting surface by the Meissner effect possesses cylindrical symmetry and thus conserves the angular momentum component along the direction $\Z$ perpedicular to the superconducting surface.  If such a levitated ferromagnet experiences a torque that would cause it to precess, in order to conserve angular momentum along $z$, the ferromagnet must tilt such that its spin component $S_z$, counteracts the rotational angular momentum component $L_z$ induced by the precession.  This in turn tilts the image dipole in such a way as to suppress the torque experienced by the ferromagnet, thereby suppressing the precession frequency.  This effect can suppress the precession frequency by orders of magnitude for ferromagnets with characteristic sizes above $\sim 0.1$\,microns.}
Alternatively, it has been proposed to place the ferromagnet in freefall~\cite{gps2021}, in order to avoid the effects of any trapping potential.

One particularly interesting application of levitated ferromagnets is the search for ultralight dark matter candidates, including axion and axion-like dark matter (henceforth, simply axion DM) and dark-photon dark matter (DPDM).  The former can address the strong-CP problem~\cite{Peccei:1977hh,Weinberg:1977ma,Wilczek:1977pj}, while both can exhibit the correct relic abundance~\cite{Preskill:1982cy,Abbott:1982af,Dine:1982ah,Nelson:2011sf,Graham:2015rva} and generically arise in new physics theories of many different origins, see, e.g., a recent review~\cite{Antypas:2022asj}.  In the ultralight regime, these candidates behave as classical fields, which oscillate near their Compton frequencies~\cite{Arias:2012az,kimball2022search}.  Axion DM could potentially couple to electron spins, causing them to precess, as if it were an AC magnetic field~\cite{terrano2019,Flower_2019,Lee2023,quax}.  As a ferromagnet is composed of many polarized electrons, axion DM can impart a collective oscillating torque on the whole ferromagnet.  In addition, axion DM and DPDM can both couple to photons, generating a physical AC magnetic field, which could also impart a torque on a ferromagnet.

While the response of a ferromagnet to an applied DC magnetic field has been well studied, both in freefall and above a superconductor~\cite{jacksonkimball2016,fadeev2021}, the response to a driving AC magnetic field has not been adequately addressed thus far.

The purpose of this work is to study the dynamics of a ferromagnet in response to an AC magnetic field and to apply these dynamics to the case of ultralight DM.  In \secref{ferromagnet}, we derive the response of the system to an applied AC magnetic field, which is qualitatively different from the DC case.  This is because, in the DC case, the precession frequency is given by the Larmor frequency, which is proportional to the applied magnetic field.  Meanwhile, we will see that in the AC case, the frequency of the ferromagnet dynamics is determined by the frequency of the AC field (the Compton frequency, in the case of ultralight DM).  Therefore, whether the ferromagnet precesses as a gyroscope or librates as a compass needle will be frequency-dependent.  Moreover, as mentioned above, the presence of a levitation/trapping mechanism can alter the dynamics of the system.  We will determine in what contexts the ferromagnet undergoes precession vs. libration.

In \secref{sensitivity}, we compute the sensitivity of a ferromagnet to an applied AC magnetic field.  We review the relevant noise sources and utilize the results of \secref{ferromagnet} to determine the magnetic-field sensitivity of a ferromagnet setup, accounting for motion in both angular directions.  We propose three cases of interest: one representative of an existing levitated setup~\cite{ahrens2024}, a future levitated setup, and a future freefall setup.  The parameter choices for these setups are shown in \tabref{parameters} and their magnetic-field sensitivities are computed in \figref{sensitivities}.

In \secref{DM}, we project the sensitivities of these setups to ultralight DM.  We review the physics of axion DM coupled to electrons, kinetically mixed DPDM, and axion DM coupled to photons.  In each case, we compute the effective/physical AC magnetic field generated by the DM candidate, and show the sensitivities of the three setups of interest to the DM candidate in \figref{projections}.

In \secref{conclusion}, we conclude.  We make all the code used in this work publicly available on Github~\cite{github}.


\section{Levitated ferromagnets}
\label{sec:ferromagnet}

In this section, we compute the response of a levitated ferromagnet to an applied AC magnetic field.  Importantly, we account for the effect of any trapping potential on the ferromagnet's response and determine when libration versus precession occurs.  We begin by introducing some examples of trapping potentials.  Then, we derive the equations of motion for the dynamics of the ferromagnet in this trap.  Finally, we show how these dynamics are modified in the presence of a driving field.

\subsection{Trapping potential}
\label{sec:potential}
Generically, in order for the ferromagnet to remain levitated, it must be trapped in both the translational and angular directions.  In other words, it must sit at the minimum $(\bm x_0,\N_0)$ of some potential $V(\bm x,\N)$.  Here $\bm x$ denotes the position of the ferromagnet, while $\N=(\theta,\phi)$ describes its orientation (using spherical coordinates with $\theta=0$ the positive $z$-axis).  In this work, we will consider the magnetic moment of the ferromagnet to be locked to its spatial orientation so that $\N$ more specifically denotes the direction of its magnetic moment.%
\footnote{In general, the individual electron spins $\bm S_i$ within the ferromagnet are not locked to its orientation $\N$.  The atomic lattice of the ferromagnet exhibits some interaction with each electron spin, which relaxes the spins to align with the lattice.  This relaxation occurs at a typical rate $\Gamma\sim\mathrm{GHz}$~\cite{bhagat}.  In this work, we only consider dynamics at much lower frequencies than this (see \citeR{Flower_2019} for an example at higher frequencies, where such spin excitations occur), and so it is safe to treat the macroscopic magnetic moment of the ferromagnet to be locked to its orientation.  We do note that coupling of individual spin fluctuations to external magnetic fields can act as an additional noise source, though this noise is typically small~\cite{Band_2018}.}
If the ferromagnet consists of $N$ polarized electron spins, then its magnetic moment is given by
\begin{equation}
    \bm\mu=-\gamma_e\bm S\equiv-\gamma_e\cdot\frac{N\hbar}2\N,
\end{equation}
where $\gamma_e=g_e\mu_B/\hbar$ is the electron gyromagnetic ratio. ($g_e$ is the electron $g$-factor and $\mu_B=e\hbar/2m_e$ is the Bohr magneton.)

There are various ways in which the ferromagnet can be trapped in a potential.  Perhaps the simplest is to levitate the ferromagnet in some static magnetic field $\bm B(\bm x)$.  This gives the trapping potential
\begin{equation}\label{eq:Btrap}
    V(\bm x,\N)=-\bm\mu\cdot\bm B(\bm x)+mgz,
\end{equation}
where the latter term arises due to gravity ($m$ is the mass of the ferromagnet, and $g$ is the gravitational accleration on Earth).  In such a potential, the ferromagnet will always prefer to align with the local magnetic field, i.e. $\N_0=\hat{\bm B}(\bm x_0)$.%
\footnote{By Earnshaw's theorem, a static magnetic field alone cannot stably levitate a magnetic dipole~\cite{Earnshaw}.  The equilibrium $\bm x_0$ can be made stable through an active feedback loop, a method known as electromagnetic levitation~\cite{moon2004superconducting}.  In this work, we focus primarily on the angular motion of the ferromagnet, and so do not worry about the stability of the translational modes.}
Note that in this case, the ferromagnet will, in general, be trapped in both angular directions, i.e. $\left.\partial^2_\theta V,\partial^2_\phi V\right|_{(\bm x_0,\N_0)}>0$.%
\footnote{We note that this spherical coordinate system becomes pathological when $\N_0=\Z$ because $\phi$ is not well-defined at this point.  In the sections that follow, we will consider only the angular dependence of $V(\N)$, in which case we will be free to rotate our coordinate system so that $\N_0\neq\Z$.}

Alternatively, the ferromagnet may be levitated above a superconducting plane.  The potential in such a setup can be computed via the method of images; that is, if the ferromagnet lies a distance $z$ above the superconducting plane, then one computes the potential it feels due to a magnetic moment located a distance $z$ below the superconducting plane~\cite{fadeev2021,vinante2021}.  This gives a potential%
\footnote{Note the additional factor of $\frac12$ in the first term of \eqref{SCtrap}.  Without this factor, this term would describe the work required to bring two physical dipoles from infinity to a distance $2z$ apart.  Because we have only one physical dipole, only half the work is required to bring it to a distance $2z$ from its image.}
\begin{align}\label{eq:SCtrap}
    V(\bm x,\N)&=-\frac12\bm\mu\cdot\frac{\mu_0}{4\pi}\frac{3(\tilde{\bm x}\cdot\tilde{\bm\mu})\tilde{\bm x}-\tilde x^2\tilde{\bm\mu}}{\tilde x^5}+mgz\\
    &=\frac{\mu_0\mu^2}{64\pi z^3}(1+\cos^2\theta)+mgz,
\label{eq:SCtrap2}\end{align}
where $\tilde{\bm x}=2z\Z$ is the distance between the ferromagnet and its image, and $\tilde{\bm\mu}=(\pi-\theta,\phi)$ is the orientation of the image magnetic moment.  It is clear in this case that the ferromagnet is trapped in the $\theta$-direction, with its minimum at $\theta_0=\pi/2$ (parallel to the superconducting plane), but it is free to rotate in the $\phi$-direction.  In a physical system, this exact degeneracy in the $\phi$-direction will be broken, but nevertheless, the trapping in the $\phi$-direction can be significantly weaker than the trapping in the $\theta$-direction, i.e. $\left.\partial^2_\theta V\right|_{(\bm x_0,\N_0)}\gg\left.\partial^2_\phi V\right|_{(\bm x_0,\N_0)}$.

\subsection{Ferromagnet dynamics}
\label{sec:dynamics}

Now let us derive the equations of motion for the ferromagnet.  We will first consider only the trapping potential, without the presence of any driving AC magnetic field.  In the remainder of this work, we will also ignore translational motion, and only focus on the angular dependence of $V(\N)$.%
\footnote{\label{ftnt:coupling}%
The translational modes of the system will not be directly excited by a uniform magnetic field but instead can only be excited by a magnetic field gradient.  The DM models of interest and the frequency regime considered in this work produce magnetic-field signals that are relatively uniform, so in this work, we will neglect any such gradients.  In principle, the translational modes may also exhibit some coupling to the angular modes, e.g. due to inhomogeneities in the trap.  This cross-coupling is small in existing experiments~\cite{vinante2020,vinante2021}.}
Let the total angular momentum of the ferromagnet be given by $\bm J=\bm S+\bm L$, which consists of both an intrinsic spin contribution $\bm S$ and an orbital angular momentum contribution $\bm L$.  The potential exerts a torque
\begin{align}\label{eq:torque}
    \bm\tau=\dd{\bm J}t&=-\N\times\nabla_{\N}V\\
    &\equiv-\N\times\left(\dd V\theta\THETA+\frac1{\sin\theta}\dd V\phi\PHI\right)
\end{align}
on the ferromagnet.  Additionally, the orientation of the ferromagnet rotates around its orbital angular momentum
\begin{equation}\label{eq:rotation}
    \dd\N t=\bm\Omega\times\N=(I^{-1}\bm L)\times\N.
\end{equation}
Generically, the moment of inertia $I$ may be an anisotropic tensor.  However, for simplicity, in this work, we will take $I$ to be diagonal, e.g., in the case of a spherical ferromagnet. 
 Note that because $\bm S\propto\N$, then we may replace $\bm L$ in this expression with $\bm J$.  If we make this replacement, then these two equations of motion govern the dynamics of $\N$ and $\bm J$.  Let us normalize all of our quantities by the intrinsic spin
\begin{align}
    \bm J&=\frac{N\hbar}2\bm j\\
    \bm L&=\frac{N\hbar}2\bm\ell\\
    \omega_I&=\frac{N\hbar}{2I}\\
    V&=\frac{N\hbar}2v,
\end{align}
so that we may rewrite the equations of motion \eqref[s]{torque} and (\ref{eq:rotation}) as
\begin{align}\label{eq:eom1}
    \dd{\bm j}t&=-\N\times\nabla_{\N}v\\
    \dd{\N}t&=\omega_I(\bm j\times\N).
\label{eq:eom2}\end{align}
The frequency $\omega_I$ is known as the Einstein-de Haas frequency~\cite{EinsteindeHaas}.

Let us decompose these equations of motion in terms of the unit vectors $\N$, $\THETA$, and $\PHI$ in spherical coordinates.  The time derivatives of these coordinates are related by
\begin{align}
    \dd\N t&=\dd\theta t\THETA+\sin\theta\dd\phi t\PHI\\
    \dd\THETA t&=-\dd\theta t\N+\cos\theta\dd\phi t\PHI\\
    \dd\PHI t&=-\sin\theta\dd\phi t\N-\cos\theta\dd\phi t\THETA,
\end{align}
and the total angular momentum $\bm j$ can be decomposed as
\begin{equation}
    \bm j=j_n\N+j_\theta\THETA+j_\phi\PHI.
\end{equation}
Note that \eqref[s]{eom1} and (\ref{eq:eom2}) imply
\begin{equation}
    \dd{j_n}t=\bm j\cdot\dd\N t+\dd{\bm j}t\cdot\N=0,
\end{equation}
so that $j_n$ is a constant of motion. If the ferromagnet is not spinning around its magnetic moment axis, then the orbital angular momentum $\bm\ell$ has no component along $\N$, and so $j_n=1$.  In the interest of maintaining generality, we will leave our results in terms of $j_n$.  As we will see in \secref{response}, the limit $j_n\rightarrow0$ will correspond to the ``compass" behavior where the ferromagnet’s angular momentum is dominated by its orbital angular momentum, while the limit $j_n\rightarrow\infty$ will correspond to the ``electron spin" behavior where it is dominated by its intrinsic angular momentum.

In terms of our spherical-coordinate variables, \eqref[s]{eom1} and (\ref{eq:eom2}) become
\begin{align}
    \dd{j_\theta}t-j_\phi\cos\theta\dd\phi t+j_n\dd\theta t&=\frac1{\sin\theta}\dd v\phi\\
    \dd{j_\phi}t+j_\theta\cos\theta\dd\phi t+j_n\sin\theta\dd\phi t&=-\dd v\theta\\
    \dd\theta t&=\omega_Ij_\phi\\
    \sin\theta\dd\phi t&=-\omega_Ij_\theta.
\end{align}
These can then be combined to give
\begin{align}\label{eq:theta_eom}
    \ddd\theta t-\frac{\sin2\theta}2\left(\dd\phi t\right)^2+j_n\omega_I\sin\theta\dd\phi t+\omega_I\dd v\theta&=0\\
    \sin^2\theta\ddd\phi t+\sin2\theta\dd\phi t\dd\theta t-j_n\omega_I\sin\theta\dd\theta t+\omega_I\dd v\phi&=0.
\label{eq:phi_eom}\end{align}
Finally, let us suppose that the motion of the ferromagnet is small,%
\footnote{In the weakly coupled limit, the dark matter signal we consider only generates a small, oscillating perturbation to the system.}
so that we may perturb \eqref[s]{theta_eom} and (\ref{eq:phi_eom}) around the minimum $\N_0=(\theta_0,\phi_0)$ of $v$. 
 Namely, let us write $\theta=\theta_0+\delta\theta$ and $\phi=\phi_0+\delta\phi$.  Then to first order, these equations become
\begin{widetext}\begin{equation}\label{eq:matrix_eom}
    \left[\partial_t^2\begin{pmatrix}1&0\\0&\sin^2\theta_0\end{pmatrix}+j_n\omega_I\sin\theta_0\partial_t\begin{pmatrix}0&1\\-1&0\end{pmatrix}+\omega_I\begin{pmatrix}v_{\theta\theta}&v_{\theta\phi}\\v_{\phi\theta}&v_{\phi\phi}\end{pmatrix}\right]\begin{pmatrix}\delta\theta\\\delta\phi\end{pmatrix}=0,
\end{equation}\end{widetext}
where $v_{\alpha\beta}=\left.\partial_\alpha\partial_\beta v\right|_{(\theta_0,\phi_0)}$.

\subsection{Response to AC magnetic field}
\label{sec:response}

\eqref{matrix_eom} encodes all the important dynamics of the ferromagnet.  To understand the structure of this equation, let us consider the effect of an AC magnetic field $\bm B(t)=\bm B_0\cos\omega t$ on the ferromagnet.%
\footnote{\label{ftnt:linear_pol}%
In this subsection, we will assume that the AC magnetic field is linearly polarized.  This is because we will be interested in distinguishing whether the ferromagnet librates or precesses, and such a notion is only well-defined if the direction of the field is fixed.  This will be the case for the magnetic-field signal from an axion-photon coupling $\bm B_{a\gamma}$ (see \secref{axion-photon}), since its direction remains fixed and only its phase oscillates.  This will, however, not be the case for an axion-wind or DPDM signal (see \secref[s]{axion-electron} and \ref{sec:dark-photon}), as the components of the axion gradient or dark photon may have different phases [e.g., see \eqref{DPDM_wave}], and so the resulting magnetic-field signal may be elliptically polarized.}
This magnetic field will have the same effect as a (time-dependent) potential of the form of the first term in \eqref{Btrap}.  If we write
\begin{equation}
    \bm B_0=B_0\bm{\hat b}=B_0\left(b_n\N_0+b_\theta\THETA_0+b_\phi\PHI_0\right)
\end{equation}
in terms of the spherical-coordinate unit vectors at $(\theta_0,\phi_0)$, then this corresponds to a normalized potential
\begin{align}
    &v_B(\theta_0+\delta\theta,\phi_0+\delta\phi,t)=\omega_L\cos\omega t\left(\N\cdot\bm{\hat b}\right)\\
    &\qquad\approx\omega_L\cos\omega t\left(b_n+b_\theta\delta\theta+b_\phi\sin\theta_0\delta\phi\right)
\end{align}
where $\omega_L=\gamma_eB_0$.  Plugging $v_B$ into \eqref[s]{theta_eom} and (\ref{eq:phi_eom}) as an additional contribution to $v$, we find that it acts as an AC driving force for the system (because $\partial_\theta v_B,\partial_\phi v_B\neq0$).  In particular, it will appear on the right-hand side of \eqref{matrix_eom} as
\begin{equation}
    -\omega_I\omega_L\cos\omega t\begin{pmatrix}b_\theta\\b_\phi\sin\theta_0\end{pmatrix}.
\end{equation}
We can then readily interpret the structure of \eqref{matrix_eom}.  The first row represents how the system responds to an applied magnetic field in the $\theta$-direction, while the second row represents how the system responds to an applied magnetic field in the $\phi$-direction.  Meanwhile, the first column represents motion in the $\theta$-direction, and the second column represents motion in the $\phi$-direction.  This tells us that the diagonal elements of \eqref{matrix_eom} indicate libration, while the off-diagonal elements indicate precession.

Let us now analyze the behavior of this system in various cases. Without loss of generality, we may orient our coordinates so that $\theta_0=\frac\pi2$, $\phi_0=0$, and $v_{\theta\phi}=v_{\phi\theta}=0$.  Moreover, let us complexify the AC magnetic field $\bm B(t)=\bm B_0e^{-i\omega t}$.  The homogeneous response of $\delta\theta$ and $\delta\phi$ will then be
\begin{equation}\label{eq:response}
    \begin{pmatrix}\delta\theta\\\delta\phi\end{pmatrix}=-\frac{N\hbar\gamma_eB_0}2\cdot\chi(\omega)\begin{pmatrix}b_\theta\\b_\phi\end{pmatrix},
\end{equation}
where the mechanical susceptibility $\chi(\omega)$ is given by
\begin{widetext}\begin{equation}\label{eq:chi}
    \chi(\omega)^{-1}=I\left[-\omega^2\begin{pmatrix}1&0\\0&1\end{pmatrix}-ij_n\omega_I\omega\begin{pmatrix}0&1\\-1&0\end{pmatrix}+\omega_I\begin{pmatrix}v_{\theta\theta}&0\\0&v_{\phi\phi}\end{pmatrix}\right].
\end{equation}\end{widetext}

\begin{figure*}[t]
\includegraphics[width=0.49\textwidth]{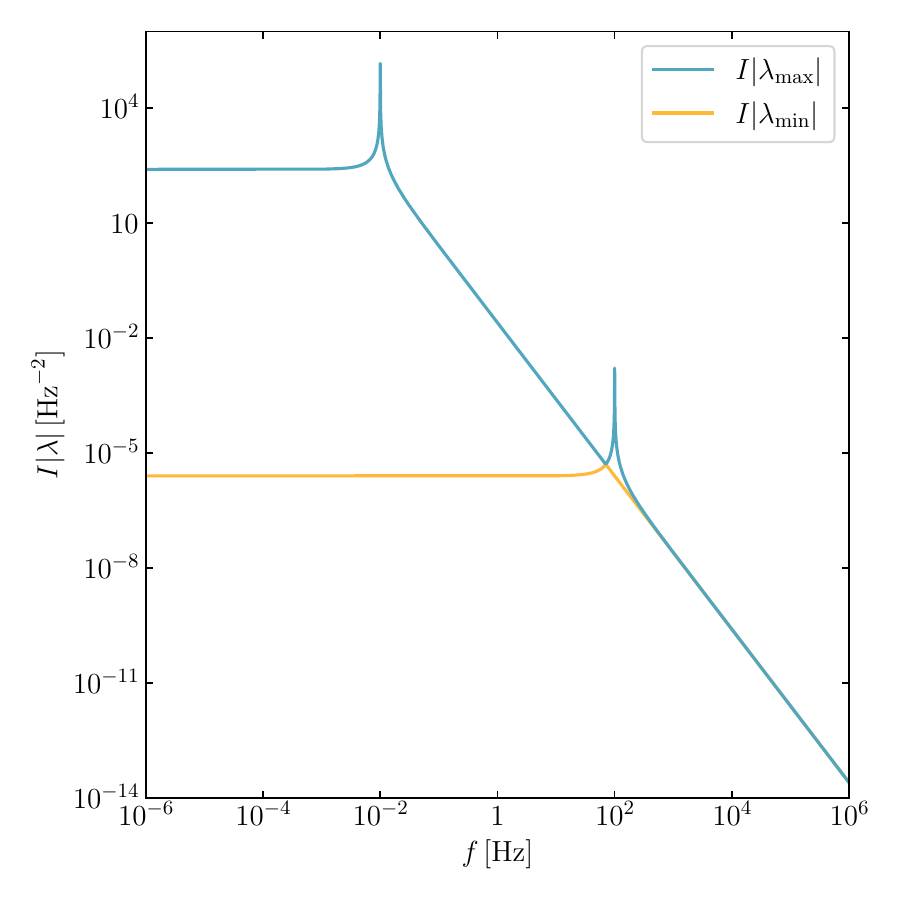}
\includegraphics[width=0.49\textwidth]{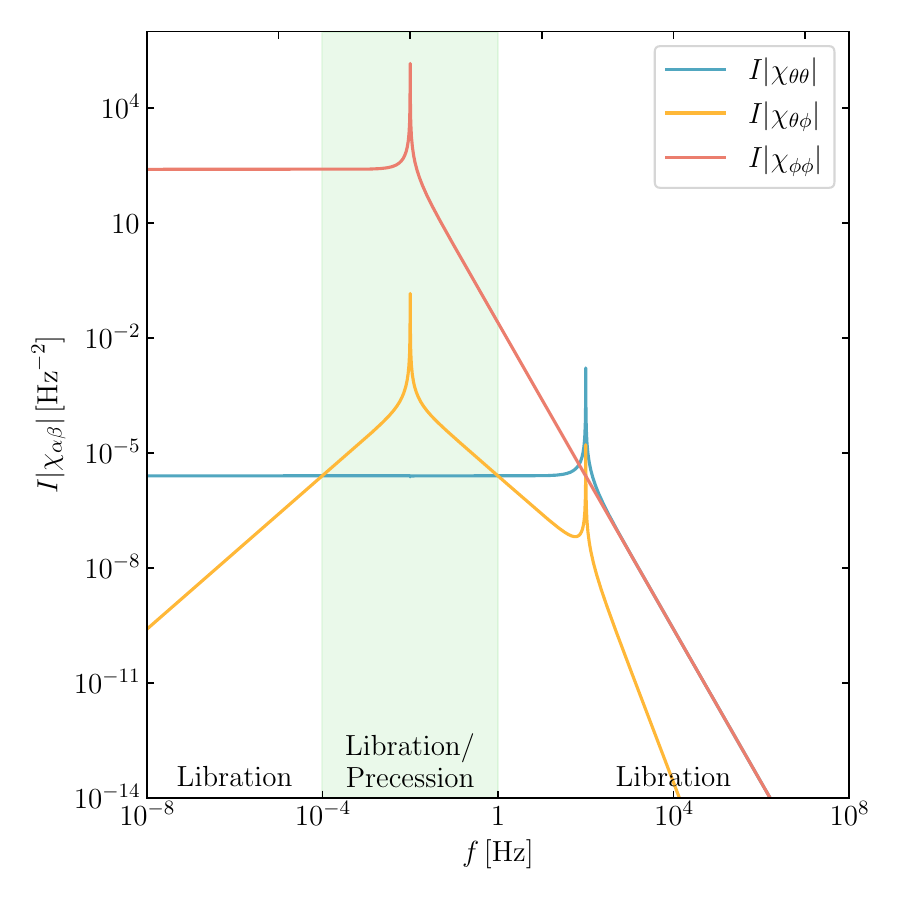}
\caption{\label{fig:trapped}%
    Absolute values of the eigenvalues (left) and elements (right) of $\chi(\omega)$ in the ``partially trapped" case.  In these plots, we set $v_{\theta\theta}=2\pi\cdot10^4\,\mathrm{Hz}$, $\omega_I=2\pi\cdot1\,\mathrm{Hz}$, $v_{\phi\phi}=2\pi\cdot10^{-4}\,\mathrm{Hz}$, and $j_n=1$.  On the left, the blue line denotes the larger eigenvalue, which predominantly determines the sensitivity of the system, while the orange line denotes the smaller eigenvalue.  Note that the larger eigenvalue exhibits resonances at $\omega=\sqrt{\omega_Iv_{\phi\phi}}$ and $\omega=\sqrt{\omega_Iv_{\theta\theta}}$.  On the right, the blue, orange, and red lines represent $I|\chi_{\theta\theta}|$, $I|\chi_{\theta\phi}|$, and $I|\chi_{\phi\phi}|$, respectively.  Note that $|\chi_{\theta\phi}|>|\chi_{\theta\theta}|$ for $j_n\omega_I\gg\omega\gg v_{\phi\phi}/j_n$ (green shaded region), indicating that an AC magnetic field in the $\theta$-direction can induce precession in this frequency range.}
\end{figure*}

\begin{figure*}[hbpt]
\includegraphics[width=0.49\textwidth]{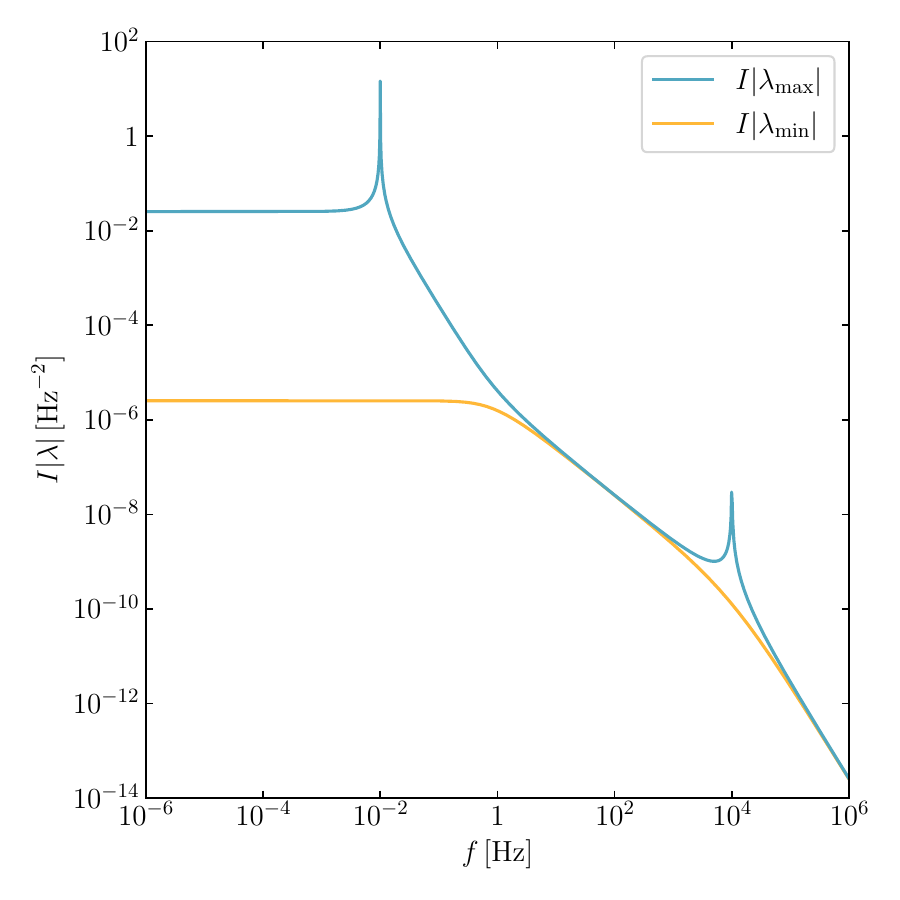}
\includegraphics[width=0.49\textwidth]{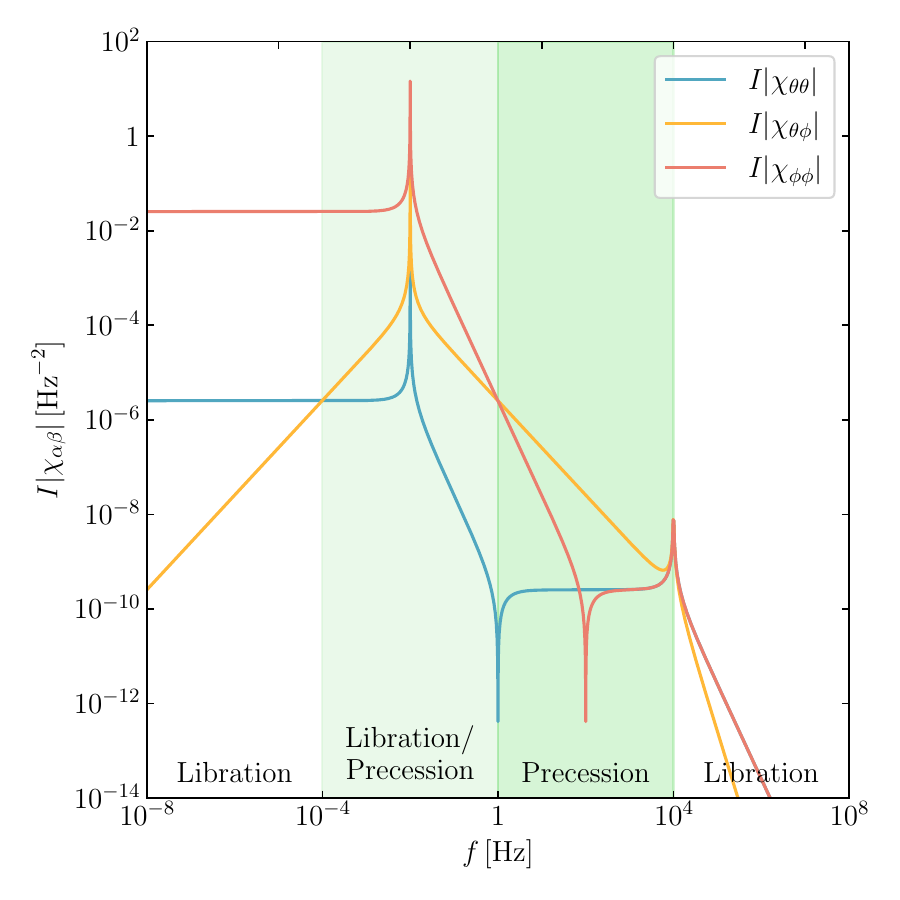}
\caption{\label{fig:gyro}%
    Same as \figref{trapped}, but in the ``gyroscope" case.  In these plots, we set $\omega_I=2\pi\cdot10^4\,\mathrm{Hz}$, $v_{\theta\theta}=2\pi\cdot1\,\mathrm{Hz}$, $v_{\phi\phi}=2\pi\cdot10^{-4}\,\mathrm{Hz}$, and $j_n=1$.  Note that the resonances in the left plot are now at $\omega=\sqrt{v_{\theta\theta}v_{\phi\phi}}$ and $\omega=\omega_I$, and that the blue line exhibits the scaling behavior indicated in \eqref{gyro_lambda}.  On the right, note that $|\chi_{\theta\phi}|>|\chi_{\theta\theta}|,|\chi_{\phi\phi}|$ for $j_n\omega_I\gg\omega\gg v_{\theta\theta}/j_n$ (darker green shaded region), indicating that there is precession in both directions in this frequency range.}
\end{figure*}

We will consider a few properties of the system based on the characteristics of $\chi(\omega)$.  First, we will be interested in the resonances of the system, which occur at the frequencies where $\chi(\omega)^{-1}$ becomes singular.  Second, we will determine whether libration or precession dominates the motion, based on whether the off-diagonal components of $\chi$ are larger than its diagonal components.%
\footnote{\label{ftnt:libration}%
One may wonder whether this definition of libration/precession is coordinate-dependent.  Because we have assumed $\bm B(t)$ is linearly polarized (see footnote~\ref{ftnt:linear_pol}), then $b_\theta$ and $b_\phi$ are real, and so we should restrict our coordinate transformations to be orthogonal (as opposed to unitary).  Given any $2\times2$ Hermitian matrix, an orthogonal transformation can always be performed so that the off-diagonal components become purely imaginary.  This coordinate choice minimizes the size of the off-diagonal components, and libration/precession can always be defined in these coordinates.  The matrix in \eqref{chi} is conveniently already in these coordinates, so we require no transformation.}
Finally, we will consider the behavior of the sensitivity as a function of frequency $\omega$.  As shown in \appref{SNR}, the peak sensitivity of the system is primarily determined by the eigenvalue of $\chi$ with the largest absolute value $\lambda_{\max}(\omega)$.  We consider three cases of interest for this system (without loss of generality, we take $v_{\theta\theta}\gg v_{\phi\phi}$ in all three cases, but their roles will simply be interchanged if the hierarchy is flipped):

\begin{itemize}
\item \textbf{Trapped} ($v_{\theta\theta}/j_n\gg v_{\phi\phi}/j_n\gg j_n\omega_I$): This will be the case, for instance, when the ferromagnet is trapped by a strong magnetic field, as in \eqref{Btrap}.  At all frequencies, the second term in \eqref{chi} can be neglected, and so the dominant motion is libration.  The system exhibits two resonances at $\omega\approx\sqrt{\omega_Iv_{\theta\theta}}$ and $\omega\approx\sqrt{\omega_Iv_{\phi\phi}}$.  For $\omega\ll\sqrt{\omega_Iv_{\phi\phi}}$, the response of the system is flat as a function of $\omega$, that is $\lambda_{\max}(\omega)\approx\left(I\omega_Iv_{\phi\phi}\right)^{-1}$.  Meanwhile for $\omega\gg\sqrt{\omega_Iv_{\phi\phi}}$, it decays as $|\lambda_{\max}(\omega)|\approx I^{-1}\omega^{-2}$ (except near the resonance $\omega\approx\sqrt{\omega_Iv_{\theta\theta}}$).

\item \textbf{Partially trapped} ($v_{\theta\theta}/j_n\gg j_n\omega_I\gg v_{\phi\phi}/j_n$): This will be the case, for instance, when the ferromagnet is trapped above a superconductor, as in \eqref{SCtrap2} [and the degeneracy in the $\phi$-direction is only weakly broken].  The resonant frequencies are again $\omega\approx\sqrt{\omega_Iv_{\theta\theta}}$ and $\omega\approx\sqrt{\omega_Iv_{\phi\phi}}$.  The off-diagonal components of \eqref{chi} are always subdominant to the $\theta\theta$-component, however, for $j_n\omega_I\gg\omega\gg v_{\phi\phi}/j_n$, they are larger than the $\phi\phi$-component.  This implies that in this frequency range, an AC magnetic field in the $\phi$-direction will result in libration, while one in the $\theta$-direction will result in precession.  The behavior of $\lambda_{\max}$ is the same as in the trapped case.  We note that in the frequency range where precession can occur, the eigenvector associated with $\lambda_{\max}$ is closely aligned with the $\phi$-direction.  Therefore even though precession can be achieved, it is not the dominant behavior of the system.

\item \textbf{Gyroscope} ($j_n\omega_I\gg v_{\theta\theta}/j_n\gg v_{\phi\phi}/j_n$): This will be the case, for instance, when the ferromagnet is in (near) freefall.  The resonant frequencies are now $\omega\approx\omega_I$ and $\omega\approx\sqrt{v_{\theta\theta}v_{\phi\phi}}$.  When $j_n\omega_I\gg\omega\gg v_{\theta\theta}/j_n$, the off-diagonal components dominate, and there is precession in both directions.  When $v_{\theta\theta}/j_n\gg\omega\gg v_{\phi\phi}/j_n$, there will be precession in one direction and libration in the other.  For all other frequencies, there will only be libration.  The response of such a system as a function of frequency is given by
\begin{equation}\label{eq:gyro_lambda}
    |\lambda_{\max}(\omega)|\approx\left\{\begin{array}{cc}\left(I\omega_Iv_{\phi\phi}\right)^{-1},&\omega\ll\sqrt{v_{\theta\theta}v_{\phi\phi}}/j_n\\
    v_{\theta\theta}/(j_n^2I\omega_I\omega^2),&v_{\theta\theta}/j_n\gg\omega\gg\sqrt{v_{\theta\theta}v_{\phi\phi}}/j_n\\
    (Ij_n\omega_I\omega)^{-1},&j_n\omega_I\gg\omega\gg v_{\theta\theta}/j_n\\
    I^{-1}\omega^{-2},&\omega\gg j_n\omega_I.\end{array}\right.
\end{equation}
\end{itemize}

In \figref[s]{trapped} and~\ref{fig:gyro}, we show the behavior of $\chi$ in the ``partially trapped" and ``gyroscope" cases, respectively.  The left plots show the behavior of the eigenvalues $\lambda_{\max},\lambda_{\min}$ of $\chi(\omega)$.  The right plots show the elements of $\chi(\omega)$.  We show in green the regions where precession is possible, which occurs when the off-diagonal elements are larger than the diagonal elements.

\section{Magnetic-field sensitivity}
\label{sec:sensitivity}

In this section, we compute the sensitivity of various ferromagnet setups to an applied AC magnetic field.  First, we review the dominant noise sources present in such a setup, accounting for noise in both angular directions using the formalism developed in \secref{ferromagnet}.  Then, we consider the physical constraints of a levitated ferromagnet setup in order to determine optimal parameters for a future levitated setup.  These parameters are shown in \tabref{parameters}, along with parameters representative of an existing setup and ones for a space-based freefall setup.  Finally, we review other potential noise sources.

\subsection{Dominant noise sources}
\label{sec:sources}

Now, we characterize the relevant noise sources in our system.  The noise analysis presented in this subsection parallels the analysis in \citeR{maglev}, but we account for the motion of the ferromagnet in both angular directions.  To this end, we generalize many of the scalar quantities introduced in \citeR{maglev} to $2\times2$ matrices [as we did for the mechanical susceptibility $\chi(\omega)$ in \eqref{chi}].

We consider three primary noise sources: thermal, imprecision, and back-action noise.  Let us first begin with thermal noise.  The thermal torque noise acting on the ferromagnet is given by
\begin{align}
    S_{\tau\tau}^\mathrm{th}(\omega)&=\begin{pmatrix}S_{\tau\tau,\theta\theta}^\mathrm{th}(\omega)&S_{\tau\tau,\theta\phi}^\mathrm{th}(\omega)\\S_{\tau\tau,\phi\theta}^\mathrm{th}(\omega)&S_{\tau\tau,\phi\phi}^\mathrm{th}(\omega)\end{pmatrix}\\
    &=4k_BI\gamma T\begin{pmatrix}1&0\\0&1\end{pmatrix},
\label{eq:therm_torque}\end{align}
where $S_{\tau\tau,\alpha\beta}$ represents the cross-correlation between torque noise in the $\alpha$- and $\beta$-directions, and $\gamma$ is the dissipation rate of the system.%
\footnote{We assume that the dissipation rate is the same for both modes, e.g., in the case of damping due to gas collisions. For a more general case, one can extend this formalism directly.}
An applied magnetic field induces a torque
\begin{equation}
    \bm\tau=-\mu\N\times\bm B\,,
\end{equation}
where $\mu=N\hbar\gamma_e/2$ is the magnetic moment of the ferromagnet, or equivalently
\begin{equation}
    \begin{pmatrix}\tau_\theta\\\tau_\phi\end{pmatrix}=\mu\begin{pmatrix}0&1\\-1&0\end{pmatrix}\begin{pmatrix}B_\theta\\B_\phi\end{pmatrix}.
\end{equation}
Then the torque noise in \eqref{therm_torque} can be translated into a magnetic-field noise
\begin{align}
    S_{BB}^\mathrm{th}(\omega)&=\frac1{\mu^2}\begin{pmatrix}0&-1\\1&0\end{pmatrix}S_{\tau\tau}^\mathrm{th}(\omega)\begin{pmatrix}0&1\\-1&0\end{pmatrix}\\
    &=\frac{4k_BI\gamma T}{\mu^2}\begin{pmatrix}0&-1\\1&0\end{pmatrix}\begin{pmatrix}1&0\\0&1\end{pmatrix}\begin{pmatrix}0&1\\-1&0\end{pmatrix}\\
    &=\frac{4k_BI\gamma T}{\mu^2}\begin{pmatrix}1&0\\0&1\end{pmatrix}.
\end{align}

Imprecision and back-action are noise sources related to the readout scheme.  For concreteness, here we will consider a readout scenario that utilizes two SQUIDs (to read the two angular modes of the ferromagnet).  Each SQUID exhibits both a flux noise $S_{\varphi\varphi,j}$ and current noise $S_{JJ,j}$ (for $j=1,2$).%
\footnote{In this work, we neglect any correlations $S_{\varphi J}$ between flux and current noise.}
They can be combined to define the energy resolution $\kappa_j=\sqrt{S_{\varphi\varphi,j}S_{JJ,j}}$ of the SQUID, which is bounded below by the uncertainty relation $\kappa_j\geq\hbar$~\cite{Voss1981}.  Currents in the SQUIDs lead to back-action torques on the ferromagnet.  These can be defined by a coupling matrix
\begin{align}
    \begin{pmatrix}\tau_\theta\\\tau_\phi\end{pmatrix}=\bm\tau&=-\N\times\eta\bm J\\
    &=\begin{pmatrix}0&1\\-1&0\end{pmatrix}\begin{pmatrix}\eta_{\theta1}&\eta_{\theta2}\\\eta_{\phi1}&\eta_{\phi2}\end{pmatrix}\begin{pmatrix}J_1\\J_2\end{pmatrix}.
\end{align}
(Note that if we wish to reduce to the case of a single-SQUID readout, this can be done by taking $\eta_{\theta2},\eta_{\phi2}\rightarrow0$.)  Likewise, fluxes in the SQUIDs correspond to angular displacements $\N=\left(\eta^{-1}\right)^T\bm\varphi$.  We can then express the current and flux noise as torque and angular uncertainties, respectively
\begin{align}
    S_{\tau\tau}^\mathrm{back}&=\begin{pmatrix}0&1\\-1&0\end{pmatrix}\eta S_{JJ}\eta^T\begin{pmatrix}0&-1\\1&0\end{pmatrix}\\
    &=\begin{pmatrix}0&1\\-1&0\end{pmatrix}\eta\begin{pmatrix}S_{JJ,1}&0\\0&S_{JJ,2}\end{pmatrix}\eta^T\begin{pmatrix}0&-1\\1&0\end{pmatrix}\\
    S_{\hat n\hat n}^\mathrm{imp}&=\left(\eta^{-1}\right)^TS_{\varphi\varphi}\eta^{-1}\\
    &=\left(\eta^{-1}\right)^T\begin{pmatrix}S_{\varphi\varphi,1}&0\\0&S_{\varphi\varphi,2}\end{pmatrix}\eta^{-1}.
\end{align}
Much like the thermal noise, it is straightforward to translate the back-action noise into a magnetic-field noise
\begin{equation}\label{eq:back_noise}
    S_{BB}^\mathrm{back}(\omega)=\frac1{\mu^2}\eta S_{JJ}\eta^T.
\end{equation}
The imprecision noise, on the other hand, requires the use of $\chi(\omega)$, as in \eqref{response}, in order to translate it into a magnetic-field noise%
\footnote{One may consider adding a damping term (corresponding to the quality factor of the system) to the definition of $\chi(\omega)$ in \eqref{chi}, in order to regulate the behavior of this expression near the resonances of $\chi(\omega)$.  Below, we will consider parameters such that imprecision noise never dominates on-resonance, so it is reasonable to exclude this damping term.}
\begin{equation}\label{eq:imp_noise}
    S_{BB}^\mathrm{imp}(\omega)=\frac1{\mu^2}\chi(\omega)^{-1}\left(\eta^{-1}\right)^TS_{\varphi\varphi}\eta^{-1}\chi(\omega)^{-1}.
\end{equation}

The total magnetic-field noise will be given by
\begin{equation}
    S_{BB}^\mathrm{tot}(\omega)=S_{BB}^\mathrm{th}(\omega)+S_{BB}^\mathrm{imp}(\omega)+S_{BB}^\mathrm{back}(\omega).
\end{equation}
As the imprecision and back-action noise scale in opposite ways with the coupling $\eta$, there exists a trade-off between them, and so we should consider our choice of $\eta$ carefully.  Let us begin by making a slight change of variables to \eqref[s]{back_noise} and (\ref{eq:imp_noise}); that is, let us define
\begin{align}
    \kappa&=S_{JJ}^{1/2}S_{\varphi\varphi}^{1/2}=\begin{pmatrix}\kappa_1&0\\0&\kappa_2\end{pmatrix}\\
    \tilde\eta&=\eta S_{JJ}^{1/4}S_{\varphi\varphi}^{-1/4},
\end{align}
so that we may write
\begin{align}
    S_{BB}^\mathrm{back}(\omega)&=\frac1{\mu^2}\tilde\eta\kappa\tilde\eta^T\\
    S_{BB}^\mathrm{imp}(\omega)&=\frac1{\mu^2}\chi(\omega)^{-1}\left(\tilde\eta^{-1}\right)^T\kappa\tilde\eta^{-1}\chi(\omega)^{-1}.
\end{align}
As we can see from \figref[s]{trapped} and~\ref{fig:gyro}, the response of the system is maximized for frequencies at/below the lowest resonance, and so this is where we will get the best sensitivity.  As such, we will choose $\eta$ to maximize our sensitivity in this region.  Note that in this frequency range, the last term in \eqref{chi} always dominates (regardless of what parameter regime we are in).  Therefore, $\chi(\omega)$ is always nearly diagonal.  Since $\kappa$ is also diagonal, it will be advantageous for us to take $\tilde\eta$ diagonal as well.  In that case, we find%
\footnote{In many cases, \eqref{imp_approx} is not the correct expression for $S_{BB}^\mathrm{imp}$, as we have neglected the contributions from the off-diagonal elements of $\chi(\omega)^{-1}$.  Nevertheless, \eqref{imp_approx} possesses the correct eigenvalues and eigenvectors for $S_{BB}^\mathrm{imp}$, which are the only properties we require.  This is because \eqref{imp_approx} has the correct value for $S_{BB,\theta\theta}^\mathrm{imp}$, which is much larger than the other elements.  This ensures that $\kappa_1\tilde\eta_1^{-2}V_{\theta\theta}^2/\mu^2$ is indeed an eigenvalue of $S_{BB}^\mathrm{imp}$, with corresponding eigenvector approximately equal to $\THETA$.  The other eigenvector is fixed by orthogonality, and the other eigenvalue is fixed by the determinant of $S_{BB}^\mathrm{imp}$ [which depends negligibly on the off-diagonal elements of $\chi(\omega)^{-1}$].}
\begin{align}
    S_{BB}^\mathrm{back}(\omega)&=\frac1{\mu^2}\begin{pmatrix}\kappa_1\tilde\eta_1^2&0\\0&\kappa_2\tilde\eta_2^2\end{pmatrix}\\
    S_{BB}^\mathrm{imp}(\omega)&\approx\frac1{\mu^2}\begin{pmatrix}\kappa_1\tilde\eta_1^{-2}V_{\theta\theta}^2&0\\0&\kappa_2\tilde\eta_2^{-2}V_{\phi\phi}^2\end{pmatrix},
\label{eq:imp_approx}\end{align}
where the approximation in \eqref{imp_approx} holds for $\omega\leq\sqrt{\omega_Iv_{\phi\phi}},\sqrt{v_{\theta\theta}v_{\phi\phi}}$, and we have defined $V_{\alpha\beta}=\left.\partial_\alpha\partial_\beta V\right|_{(\theta_0,\phi_0)}$ and set
\begin{equation}\label{eq:eta_diag}
    \tilde\eta=\begin{pmatrix}\tilde\eta_\theta&0\\0&\tilde\eta_\phi\end{pmatrix}.
\end{equation}

Once we have chosen $\tilde\eta$ to be diagonal, we see that the choice of coupling along each axis is independent.  As shown in \appref{SNR}, the total sensitivity of our system ultimately depends on the sensitivity in both directions, but it will be predominantly determined by the sensitivity along the more sensitive axis.  Let us first address how to choose the coupling for a single axis.  There are two cases one should consider.  First, if the thermal noise is larger than the geometric mean of the back-action and low-frequency imprecision noise, that is,
\begin{equation}
    S_{BB,\alpha\alpha}^\mathrm{th}\geq\sqrt{S_{BB,\alpha\alpha}^\mathrm{imp}(\omega=0)\cdot S_{BB,\alpha\alpha}^\mathrm{back}},
\end{equation}
or equivalently
\begin{equation}\label{eq:therm_dom}
    \tilde\eta_\alpha^{(\mathrm{res})}\geq\tilde\eta_\alpha^{(\mathrm{broad})},
\end{equation}
where
\begin{align}\label{eq:eta_res}
    \tilde\eta_\alpha^{(\mathrm{res})}&=\sqrt{\frac{4k_BI\gamma T}{\kappa_j}}\\
    \tilde\eta_\alpha^{(\mathrm{broad})}&=\sqrt{V_{\alpha\alpha}},
\label{eq:eta_broad}\end{align}
then both imprecision and back-action noise can be made subdominant to thermal noise at frequencies at/below the resonance of this mode.  This is achieved so long as
\begin{equation}\label{eq:therm_dom_eta}
    \tilde\eta_\alpha^{(\mathrm{res})}\geq\tilde\eta_\alpha\geq\left[\tilde\eta_\alpha^{(\mathrm{broad})}\right]^2/\tilde\eta_\alpha^{(\mathrm{res})}.
\end{equation}
The closer $\tilde\eta$ is to the upper bound in \eqref{therm_dom_eta}, the better the sensitivity will be at higher frequencies (as imprecision noise always dominates at sufficiently high frequencies), but if the primary goal is to maximize sensitivity at/below the resonance, then any coupling in this range will suffice.

If \eqref{therm_dom} is not met, then one of imprecision or back-action noise will always dominate at low frequencies.  There are then two possible approaches.  If we wish to maximize the sensitivity on-resonance, then we should set the back-action noise equal to thermal noise, i.e. the choice of $\tilde\eta$ in \eqref{eta_res}.  If, instead, we wish to optimize for sensitivity at low frequencies, then we should set back-action noise equal to low-frequency imprecision noise, i.e., the choice of $\tilde\eta$ in \eqref{eta_broad}.  In this way, an individual mode can be optimized for either resonant or broadband detection.  If two SQUIDs are utilized to track both modes, then we can make this choice for each mode separately.

\subsection{Parameter estimation}
\label{sec:parameters}

In this section, we estimate the parameters that can be realistically achieved in a future levitation setup.  In particular, we need estimations for the ferromagnet parameters, the trapping potential $V$, the temperature $T$ and dissipation rate $\gamma$ that determine the thermal noise, and finally, the energy resolution $\kappa$ and coupling $\tilde\eta$ of the readout.

For concreteness, we consider a setup with a permanent hard ferromagnetic sphere with magnetization $M=B_s/\mu_0$ (with $B_s$ saturation remanence field), density $\rho$, and radius $R$, levitated via the Meissner effect above a type-I superconducting plane \cite{vinante2020, vinante2021}, i.e., with a potential of the form in \eqref{SCtrap2}. We consider the plane to be made of lead, which has a critical field $B_c=80$\,mT. The equilibrium levitation height can be expressed as \cite{vinante2020}
\begin{equation}\label{eq:levheight}
    z_0 = \left( \frac{\mu_0 M^2 R^3}{16 \rho g } \right)^\frac{1}{4}.
\end{equation}

\begin{figure}[t]
\includegraphics[width=\columnwidth]{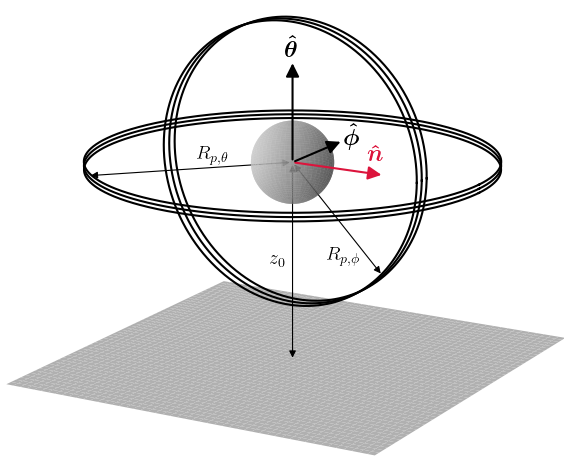}
\caption{\label{fig:pickup}
    Scheme of the circular pick-up coils considered for the estimation of coupling. The spherical magnet at equilibrium height $z_0$ has its magnetic dipole oriented along $\N$.  Rotations along the spherical angles $\theta$ and $\phi$ are detected with maximum efficiency by the horizontal coil (radius $R_{p,\theta}$) and vertical coil (radius $R_{p,\phi}$) respectively.}
\end{figure}

A typical neodymium-based rare earth alloy used in current experiments features $M\approx7\times10^5\,\mathrm{A/m}$ and $\rho \approx 7400$\,kg/m$^3$.  For these parameters, the maximum field produced by the magnet at the superconducting surface
\begin{equation}\label{eq:surface_field}
    B_{\mathrm{surf}} = \frac{2B_s}3\left(\frac R{z_0}\right)^3
\end{equation}
increases with $R$ and approaches the critical field for $R \approx 35$\,mm. This is an ultimate upper limit on $R$ for pure Meissner levitation above lead. A safe choice for a future experiment is $R = 2$ mm, which implies a field at the surface of $\sim9$\,mT, one order of magnitude below the critical field.  Levitating a larger magnet would require using a type-II superconductor \cite{gieseler2020}, in which case modeling would be more complex, and additional dissipation from vortex motion would arise.

The magnetic confinement in the polar direction $V_{\theta \theta}$ can be determined via \eqref{SCtrap2}. The azimuthal confinement $V_{\phi\phi}$ has been recently shown to be tunable in a wide range between $10^{-5}$ and $10^{-1}$ times $V_{\theta \theta}$ by applying a bias field \cite{ahrens2024}. For concreteness we set $V_{\phi\phi} = 10^{-3} V_{\theta \theta}$. For the thermal noise we set $\gamma=2\times 10^{-6}$ Hz and $T=50$ mK. Such values appear within reach and have been approached by a recent experiment~\cite{Fuchs2024}, where $\gamma \lesssim 10^{-5} $ Hz was measured at the operating temperature $T=30$ mK. In that experiment, an excess noise of a factor $100$ larger than the thermal noise was attributed to insufficient vibrational isolation. We also remark that even lower dissipation $\gamma\approx 4 \times 10^{-7}$ Hz has been measured with a nanoparticle in ultrahigh vacuum levitated within a Paul trap \cite{Dania2024}.

For the readout, we consider two circular superconducting pick-up coils.  In accordance with the choice of setting $\tilde\eta$ to be diagonal [see \eqref{eta_diag}], we optimally orient the coils to sense rotations along the $\theta$- and $\phi$-directions, with number of loops $N_{p,\alpha}$ and radius $R_{p,\alpha}$.  This arrangement is sketched in \figref{pickup}.  Each pick-up coil, with inductance $L_p$, is connected to the input coil of a DC SQUID. The latter has inductance $L_S$, the input coil has inductance $L_i$, and they have mutual inductance  $M_i=k\sqrt{L_i L_S}$, with $k \leq 1$ a geometrical coupling factor.

A DC SQUID can be modeled as a linear detector of magnetic flux, with imprecision flux noise $S_{\varphi_S\varphi_S}$ and circulating current back-action noise $S_{J_SJ_S}$.  These can alternatively be expressed as flux/current energy resolution $\kappa_\varphi = S_{\varphi_S\varphi_S}/L_S$ and $\kappa_J=S_{J_SJ_S}L_S$ (so that $\kappa^2=\kappa_\varphi\kappa_J$).%
\footnote{We note that in the literature, these energy resolutions are sometimes defined as $\epsilon_\varphi=\kappa_\varphi/2$ and $\epsilon_J=\kappa_J/2$.  Here, we omit the factor of 2 to agree with our definition of $\kappa$.}
The advantage of this normalization is that the quantum limit for each noise source is given by $\kappa_\varphi,\kappa_J\geq\hbar$.

\begin{figure}[t]
\includegraphics[width=\columnwidth]{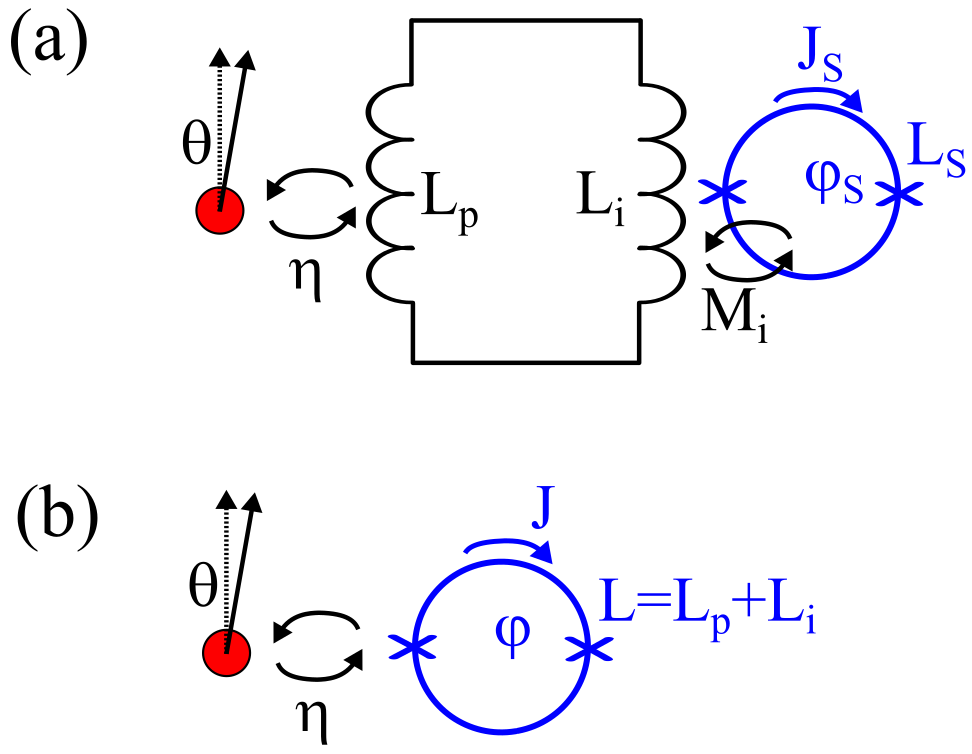}
\caption{\label{fig:SQUID}
    A SQUID connected through a superconducting pick-up coil to the ferromagnet motion [subfigure (a)] is equivalent to a SQUID directly connected to the ferromagnet [subfigure (b)], with the backward scaling $J/J_S=M_i/L$ and the forward scaling $\varphi_S/\varphi=M_i/L$. Here, $M_i$ is the mutual inductance between the input coil and SQUID, and $L=L_i+L_p$. The effective coupling $\eta$ to the equivalent SQUID coincides with the coupling to the pick-up coil. The imprecision and back-action noises $S_{\phi\phi}$ and $S_{JJ}$ of the equivalent SQUID are rescaled by $(M_i/L)^{-2}$ and $(M_i/L)^2$ with respect to the real SQUID noises.  As a result, $\kappa$ is the same for the effective and real SQUIDs, while $\tilde\eta$ is rescaled by $M_i/L$.}
\end{figure}

In a pick-up coil configuration, the flux effectively coupled into the SQUID is $\varphi_S = (M_i/L)\varphi$, where $\varphi = \eta_\theta \theta$ or $\varphi = \eta_\phi \phi$ is the flux coupled into the pick-up coil by a rotation angle $\theta$ or $\phi$ of the ferromagnet. Here, $L=L_i+L_p$ is the total inductance of the superconducting flux transformer loop. Likewise, a circulating current in the SQUID $J_S$ translates into a current $J=(M_i/L)J_S$ in the pick-up coil. This is shown in \figref{SQUID}. The flux transformer loop thus behaves as an equivalent SQUID with inductance $L$, imprecision noise $S_{\varphi \varphi} = S_{\varphi_S \varphi_S}/(M_i/L)^2$ and back-action noise $S_{JJ}=S_{J_S J_S}(M_i/L)^2$.
The energy product $\kappa=S_{JJ}S_{\varphi\varphi}$ is invariant, while $\tilde\eta=(M_i/L)\tilde\eta_S$ is rescaled.

The coupling $\eta$ between the rotating ferromagnet and pick-up coil for the optimal geometrical configuration shown in \figref{pickup} is given by
\begin{equation}
  \eta = \frac{N_p\mu_0 \mu}{2 R_p}.
\end{equation}
If one wishes to increase the coupling $\tilde\eta$, then one should maximize the product $\eta\cdot M_i/L$.  This is most easily done by modifying the pick-up coil, specifically by varying the number of loops.  A pick-up coil made of $N_p$ circular superconducting loops of radius $R$ with wire radius $a$ has inductance
\begin{equation}
   L_p=N_p^2 \mu_0 R_p \left[\log\left(\frac{8R_p}{a_p}\right)-2\right].
\end{equation}
As a function of $N_p$, the product $\eta\cdot M_i/L$ will be maximized when $L_p=L_i$.  (Alternatively, if one wishes to reduce $\tilde\eta$, the number of coils can be decreased/increased.)

\begin{table*}[t]
    \centering
    \begin{tabular}{c|c c c}
        \hline\hline
        Parameter & Existing & Future & Freefall\\ 
        \hline
        Ferromagnet radius $R$&20\,$\mu$m&2\,mm&2\,cm\\
        Ferromagnet magnetization $M$&\multicolumn{3}{c}{$7\times10^5\,\mathrm{A/m}$}\\
        Ferromagnet density $\rho$&\multicolumn{3}{c}{$7400\,\mathrm{kg/m}^3$}\\
        \hline
        Temperature $T$&4\,K&50\,mK&300\,K\\
        Dissipation rate $\gamma$&$10^{-2}\,\mathrm{Hz}$&$2\times10^{-6}\,\mathrm{Hz}$&$10^{-10}\,\mathrm{Hz}$\\
        Azimuthal trapping $V_{\phi\phi}$&$10^{-14}\,\mathrm{J}$&$10^{-3}V_{\theta\theta}$&$7\times10^{-9}\,\mathrm{J}$\\
        \hline
        Energy resolution $\kappa_\theta=\kappa_\phi$&$1000\hbar$&$\hbar$&$\hbar$\\
        Polar coupling $\tilde\eta_\theta$&$1.1\times10^{-7}\,\sqrt{\mathrm J}$&$3.7\times10^{-3}\,\sqrt{\mathrm J}$&$10^{-5}\,\sqrt{\mathrm J}$\\
        Azimuthal coupling $\tilde\eta_\phi$&$5\times10^{-9}\,\sqrt{\mathrm J}$&$3.7\times10^{-3}\,\sqrt{\mathrm J}$&$10^{-5}\,\sqrt{\mathrm J}$\\
        \hline
        $\tilde\eta_\theta^{(\mathrm{res})}=\tilde\eta_\phi^{(\mathrm{res})}$&$9.1\times10^{-7}\,\sqrt{\mathrm{J}}$&$4.6\times10^{-3}\,\sqrt{\mathrm{J}}$&$2.5\,\sqrt{\mathrm{J}}$\\
        $\tilde\eta_\theta^{(\mathrm{broad})}$&$6.4\times10^{-7}\,\sqrt{\mathrm{J}}$&$3.6\times10^{-3}\,\sqrt{\mathrm{J}}$&$10^{-5}\,\sqrt{\mathrm{J}}$\\
        $\tilde\eta_\phi^{(\mathrm{broad})}$&$10^{-7}\,\sqrt{\mathrm{J}}$&$1.1\times10^{-4}\,\sqrt{\mathrm{J}}$&$10^{-5}\,\sqrt{\mathrm{J}}$\\
        $\omega_I$&$2\pi\cdot0.53\,\mathrm{Hz}$&$2\pi\cdot5.3\times10^{-5}\,\mathrm{Hz}$&$2\pi\cdot5.3\times10^{-7}\,\mathrm{Hz}$\\
        $v_{\theta\theta}$&$2\pi\cdot4.9\times10^5\,\mathrm{Hz}$&$2\pi\cdot1.6\times10^7\,\mathrm{Hz}$&$2\pi\cdot0.12\,\mathrm{Hz}$\\
        $v_{\phi\phi}$&$2\pi\cdot1.2\times10^4\,\mathrm{Hz}$&$2\pi\cdot1.6\times10^4\,\mathrm{Hz}$&$2\pi\cdot0.12\,\mathrm{Hz}$\\
        \hline\hline
    \end{tabular}
    \caption{\label{tab:parameters}%
    Parameters choices for various setups.  Here, we show three sets of parameters: one representative of an existing levitated setup~\cite{ahrens2024} (but with an additional readout mode; see text), a future levitated setup, and a space-based freefall setup with parameters comparable to LISA Pathfinder.  Each section of the table includes ferromagnet parameters, system parameters, readout parameters, and resulting quantities defined in \secref[s]{ferromagnet} and~\ref{sec:sources}.  In the first two cases, the polar trapping $V_{\theta\theta}$ can be computed via \eqref{SCtrap2}, while in the third case, it is the same as $V_{\phi\phi}$.  In the future setup, both modes satisfy \eqref{therm_dom_eta}, so the readout is appropriately coupled.  In the existing setup, both modes are undercoupled.  In the freefall setup, they satisfy \eqref{therm_dom_eta}, but the sensitivity would benefit at higher frequencies from an even larger coupling.  In all three cases, the system exhibits ``trapped" behavior (i.e. $v_{\theta\theta},v_{\phi\phi}\gg\omega_I$).}
\end{table*}

\begin{figure*}[t]
\includegraphics[width=0.49\textwidth]{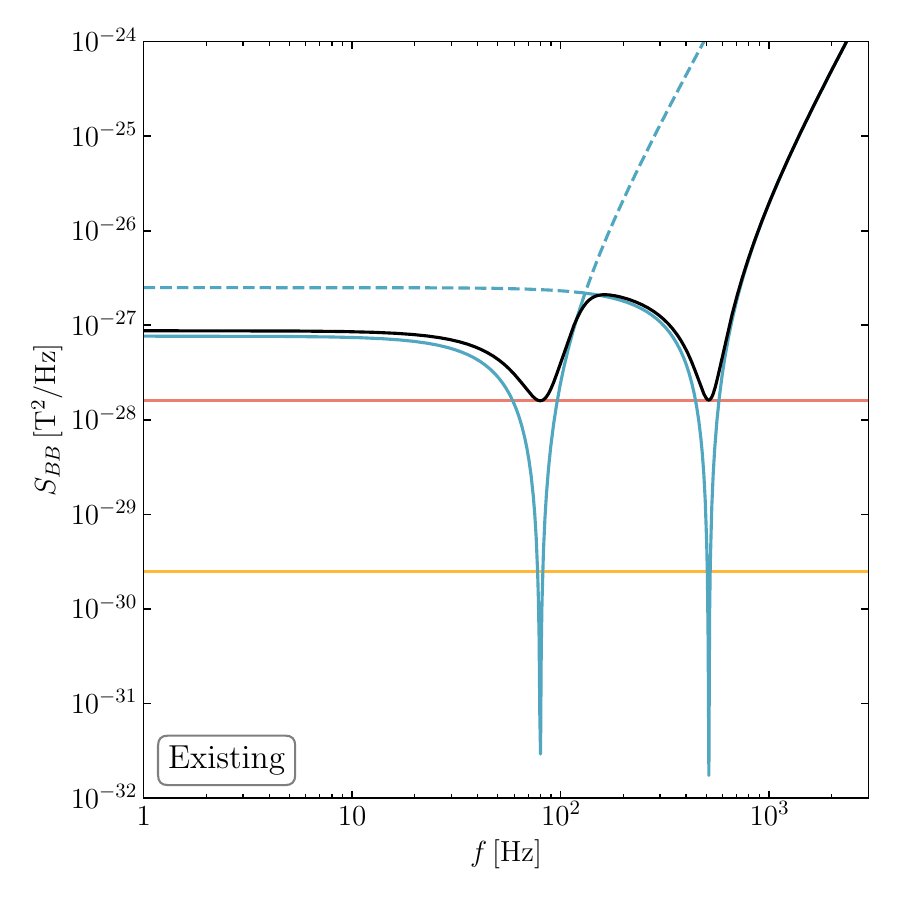}
\includegraphics[width=0.49\textwidth]{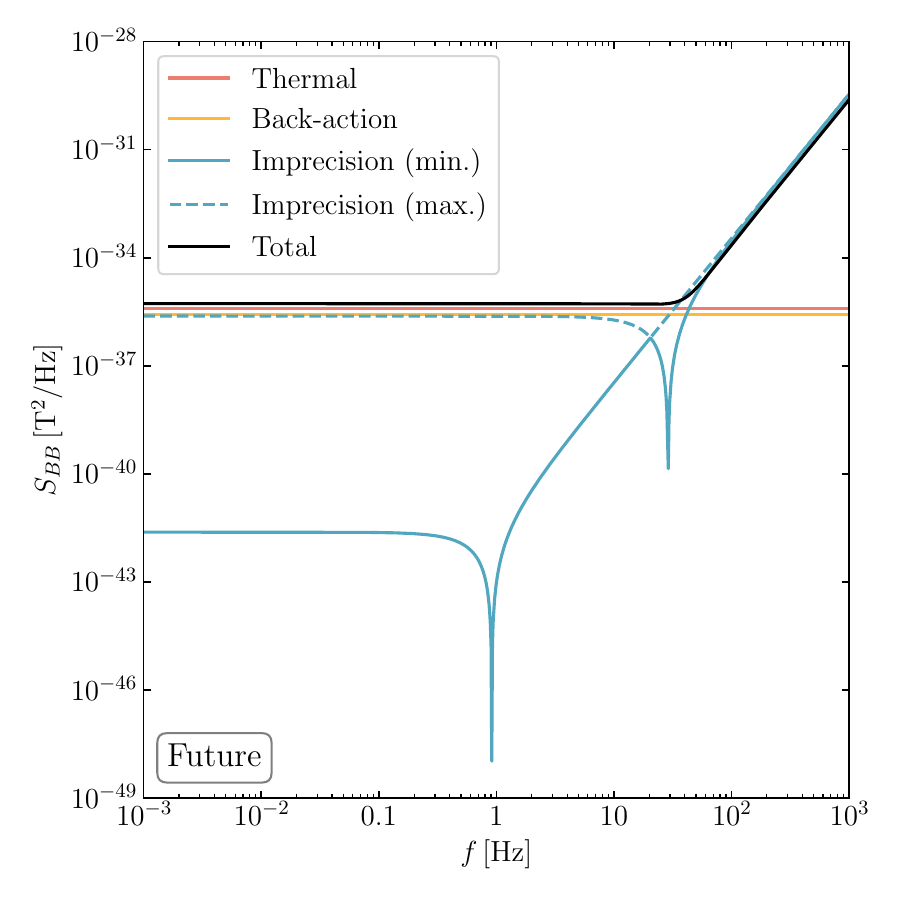}
\includegraphics[width=0.49\textwidth]{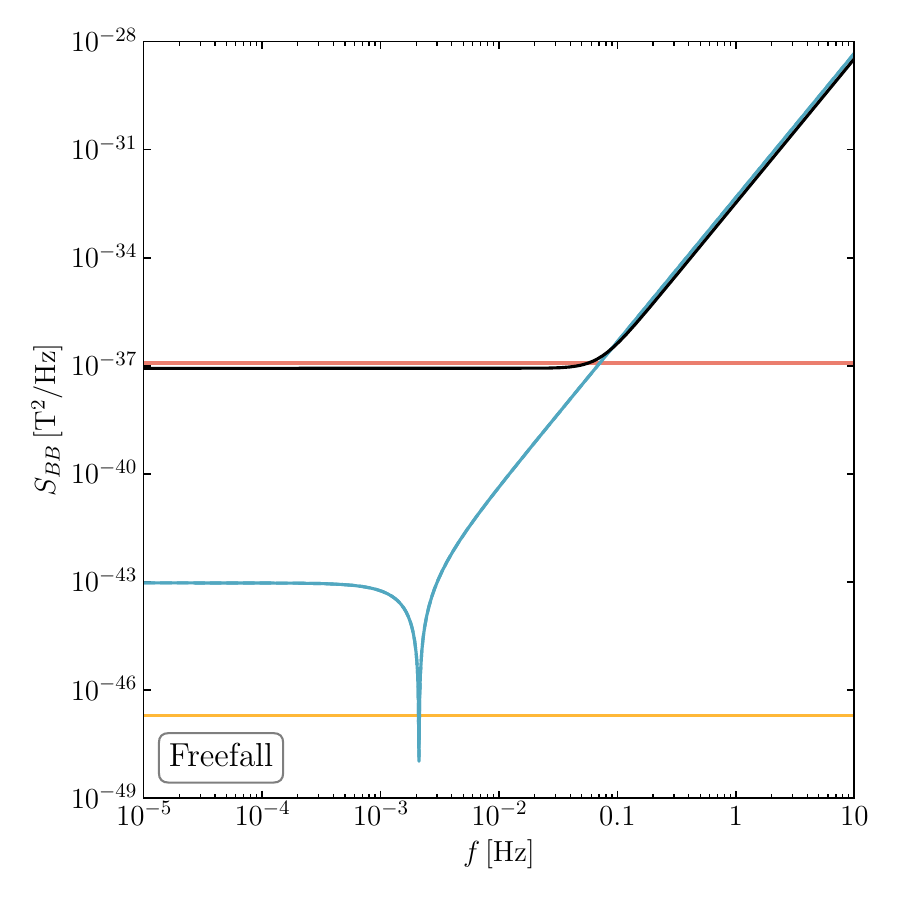}
\caption{\label{fig:sensitivities}%
    Magnetic field sensitivities for the three setups shown in \tabref{parameters}.  We show the thermal contribution in red, the backaction contribution in yellow, and the imprecision contribution in blue.  For the imprecision noise, the smaller eigenvalue of $S_{BB}^\mathrm{imp}$, which dominates the sensitivity, is shown as solid, while the larger eigenvalue is dashed.  In black, we show the total noise, specifically $\mathrm{Tr}\left[\left(S_{BB}^\mathrm{tot}\right)^{-2}\right]^{-1/2}$, as this is the quantity which appears in \eqref{SNR}.%
    \footnote{In some cases, the black curve appears lower than the colored curves by a factor of $\sqrt2$, as the colored curves show the \emph{eigenvalues} of the individual contributions.}
    Note that the future and freefall setups have broadband sensitivity because they are adequately coupled [satisfy \eqref{therm_dom_eta}], while the existing case has resonant sensitivity (in both modes) because it is undercoupled.  The freefall setup demonstrates better sensitivity than the future setup at low frequencies, but becomes dominated by imprecision noise at much lower frequencies than in the future case.  This can be improved by increasing the coupling towards the upper bound in \eqref{therm_dom_eta}.  Both cases show significantly better sensitivity than the existing case.}
\end{figure*}

In \tabref{parameters}, we show sample parameters for three different setups: ones representative of an existing levitated setup~\cite{ahrens2024}, ones for a future levitated setup, and ones for a space-based freefall setup.  Their corresponding magnetic field sensitivities $S_{BB}(f)$ are shown in \figref{sensitivities}.  The future setup takes the ferromagnet ($R$, $M$, and $\rho$) and system parameters ($T$, $\gamma$, and $V_{\phi\phi}$) described above.  For the readout, we take $R_p=8\,\mathrm{mm}$, $a_p=100\,\mu\mathrm m$, $L_S=80\,\mathrm{pH}$, $L_i=1.8\mu\mathrm H$, and $k=0.85$.  With these values, the optimal number of loops (to achieve $L_p\approx L_i$) is $N_p=6$.  We also assume a quantum-limited readout $\kappa=\kappa_\varphi=\kappa_J=\hbar$.  With these parameter choices, we see that the system exhibits the ``trapped" behavior described in \secref{response} because $v_{\theta\theta},v_{\phi\phi}\gg\omega_I$.  (We set $j_n=1$ in all cases.)  Moreover, both modes satisfy \eqref{therm_dom_eta}, so the readout is appropriately coupled.

The existing case uses ferromagnet and system parameters comparable to the setup in \citeR{ahrens2024}.  For the readout of the $\theta$-mode, we take the same parameters as in the future setup, but with a smaller pick-up loop $R_p=1\,\mathrm{mm}$ and worse energy resolution $\kappa=\kappa_\varphi=\kappa_J=1000\hbar$.  With this pick-up loop radius, the optimal number of coils is $N_p=25$.  For the readout of the $\phi$-mode, the setup in \figref{pickup} is not achievable since the equilibrium point of the ferromagnet is so close to the superconducting plane $z_0\approx250\,\mu\mathrm m$.  Instead, different geometries may be realized.  For instance, in \citeR{ahrens2024}, the $\phi$-mode is read out using a figure-eight-shaped coil lying in a plane above the ferromagnet.  Such a geometry will naturally exhibit a weaker coupling than the $\theta$-readout.  In \tabref{parameters}, we simply fix a coupling $\tilde\eta_\phi=5\times10^{-9}\,\sqrt{\mathrm J}$ comparable to that of \citeR{ahrens2024}, without focusing on any particular geometric realization.  In this case, neither mode satisfies \eqref{therm_dom_eta}, so both modes are undercoupled.

Finally, the freefall case considers a space-based experiment with parameters comparable to the LISA Pathfinder mission~\cite{LISAnanonewton,LISAbeyond}.  We take a ferromagnet of similar dimensions to the LISA test mass (but we take a neodymium sphere rather than a gold cube).  We consider the system to be at room temperature $T=300$\,K with a dissipation rate $\gamma=10^{-10}\,\mathrm{Hz}$ that produces a thermal noise slightly better than LISA Pathfinder's angular sensitivity~\cite{LISAnanonewton}.  The ferromagnet will experience some weak trapping from stray DC magnetic fields inside the apparatus.  The Sun's magnetic field near LISA's position averages $\sim5\,\mathrm{nT}$, which can be reduced by a few orders of magnitude with moderate magnetic shielding.  On the other hand, the shielding itself will exhibit some residual magnetization, which will likely dominate the stray fields inside the apparatus.  We assume a residual magnetic field of $\sim300\,\mathrm{pT}$~\cite{Ayres2024},%
\footnote{We note that it may be possible to reduce the stray fields in the apparatus further~\cite{Budker_2007,budker2013optical}.  Ultimately, trapping by stray fields does not limit our final sensitivity.  This will only affect the imprecision noise at low frequencies, but as can be seen from the bottom plot of \figref{sensitivities}, we are dominated by thermal noise at low frequencies.}
resulting in a trapping potential $V\sim7\times10^{-9}\,\mathrm{J}$.  Finally, a SQUID readout will be difficult to implement effectively at room temperature.  Instead, LISA utilizes an interferometric readout with positional imprecision noise $\sqrt{S_{\bm x\bm x}}\sim3\times10^{-14}\,\mathrm{m/\sqrt{Hz}}$~\cite{LISAbeyond}, which translates to an angular imprecision $\sqrt{S_{\N\N}}\sim10^{-12}\,\mathrm{rad/\sqrt{Hz}}$.  Assuming a quantum-limited readout $\kappa=\hbar$, this corresponds to a coupling $\tilde\eta\sim10^{-5}\,\mathrm{J}$.  We see that both modes satisfy \eqref{therm_dom_eta}, so that the low-frequency sensitivity is dominated by thermal noise.  However, since $\tilde\eta$ is far from the upper bound of \eqref{therm_dom_eta}, the high-frequency sensitivity could be improved if the coupling can be increased further.  Also note that despite the fact that the experiment is in freefall, this case still exhibits the ``trapped" rather than ``gyroscope" behavior (i.e., $v_{\theta\theta},v_{\phi\phi}\gg\omega_I$ still).  To achieve the gyroscope behavior with such a large ferromagnet would require significantly better shielding.

\subsection{Other noise sources}
\label{sec:other}

Before we move on to estimate the sensitivity of these setups to DM, we note a couple of additional sources of noise, which are not inherent but may take additional care to mitigate.  The first is vibrational noise, which can lead to translational motion of the ferromagnet if not properly attenuated.  If the translational and rotational motions of the ferromagnet exhibit some small coupling (see footnote~\ref{ftnt:coupling}), this will translate into noise in its angular orientation $\N$ (similar to imprecision noise).  The angular imprecision noise for the future levitated setup is $\sqrt{S_{\N\N}}\sim3\times10^{-15}\,\mathrm{rad}/\sqrt{\mathrm{Hz}}$.  Assuming a $\mathcal O(0.01)$ coupling between the translational and rotational modes, this setup would require a vibrational noise $\sqrt{S_{\bm x\bm x}}\lesssim6\times10^{-16}\,\mathrm m/\sqrt{\mathrm{Hz}}$ in order for vibrations to be subdominant.  The corresponding requirement for the existing and freefall setups is $\sqrt{S_{\bm x\bm x}}\lesssim10^{-12}\,\mathrm m/\sqrt{\mathrm{Hz}}$.  LIGO has achieved vibrational noises below these thresholds for frequencies $f\gtrsim10\,\mathrm{Hz}$~\cite{Nguyen_2021}.

Another noise source of concern is $1/f$ noise in the SQUID readout, which typically dominates at low frequencies $f\lesssim10\,\mathrm{kHz}$ and results in flux noise $\sqrt{S_{\varphi\varphi}}\sim5-10\,\mu\Phi_0/\sqrt{\mathrm{Hz}}$ at $f\sim1\,\mathrm{Hz}$~\cite{Sendelbach} (compared to the much lower noise $\sqrt{S_{\varphi\varphi}}\sim0.04\,\mu\Phi_0/\sqrt{\mathrm{Hz}}$ assumed in our future setup). Methods to substantially reduce $1/f$ noise by material engineering have demonstrated suppression down to $\sqrt{S_{\varphi\varphi}}\sim0.3\,\mu\Phi_0/\sqrt{\mathrm{Hz}}$ at $f\sim1\,\mathrm{Hz}$~\cite{Sendelbach}. A more complex approach is to mitigate $1/f$ noise by upconverting the signal to a higher pump frequency where $1/f$ noise is negligible. This can be achieved via a capacitor bridge transducer~\cite{Cinquegrana1993, Paik2016}, or an inductance bridge transducer~\cite{Paik1986}.  Henceforth, we assume that our readout implements such a scheme. This allows us to neglect $1/f$ noise and validates our assumption of frequency-independent flux noise.

\section{Searching for ultralight DM}
\label{sec:DM}

In this section, we introduce a few ultralight DM candidates/couplings which could be detected with ferromagnets.  All of these candidates manifest in laboratory experiments as effective/physical AC magnetic fields, and so a ferromagnet will respond to these DM candidates in the manner described in \secref{ferromagnet}.  We can then use the magnetic-field sensitivities computed in \secref{sensitivity} to project the sensitivity of ferromagnets to these DM candidates.  In this section, we consider two possible interactions of axion DM: a coupling to electrons $g_{ae}$ and a coupling to photons $g_{a\gamma}$.  The former directly causes precession of electron spins, while the latter creates an observable magnetic field (which in turn leads to precession of magnetic moments).  In addition, we also consider DPDM with kinetic mixing $\varepsilon$, which generates a similar observable magnetic field.

\subsection{Axion-electron coupling}
\label{sec:axion-electron}

An axionlike particle $a$, with mass $m_a$, is a pseudoscalar which may generically exhibit various interactions with SM particles.  One such possible interaction is a coupling to electrons via the operator
\begin{equation}\label{eq:axion_electron}
    \LL_{ae}\supset\frac{g_{ae}\sqrt{\hbar^3c}}{2m_e}\partial_\mu a\bar\psi_e\gamma^\mu\gamma_5\psi_e,
\end{equation}
where $m_e$ is the electron mass and $\psi_e$ is its wavefunction.  When the electron is non-relativistic, this leads to a coupling between the axion gradient and electron spins, i.e. a Hamiltonian of the form
\begin{equation}
    H\supset\frac{g_{ae}\sqrt{\hbar^3c}}{2m_e}\bm\sigma_e\cdot\nabla a\equiv-\gamma_e\bm S_e\cdot\bm B_{ae},
\end{equation}
where $\bm S_e=\frac\hbar2\bm\sigma_e$ is the spin of the electron (and $-\gamma_e\bm S_e$ is its magnetic moment).  We then see that an axion gradient (or ``axion wind") has the same effect on an electron spin as an effective magnetic field%
\footnote{We note that $\bm B_{ae}$ may receive a suppression from the magnetic shielding of the experimental apparatus in certain contexts [analogous to the $m_{A'}L$ shielding suppression appearing in the DPDM signal in \eqref{DPDM_signal}].  In particular, this can occur if the shielding is accomplished with a material of high permeability, such as mu-metal, but it will not occur if superconducting shielding is used~\cite{jacksonkimball_shielding}.  Note that in the case of multiple layers of shielding, it is only the composition of the innermost layer which is relevant for this suppression.}
\begin{equation}\label{eq:Baxion}
    \bm B_{ae}=-\frac{2g_{ae}\sqrt{\hbar c}}{g_ee}\nabla a.
\end{equation}
As a ferromagnet is composed of many polarized electrons spins, the axion wind will also generate a torque on the ferromagnet, just as a real magnetic field would.  (See also \citeR{Flower_2019} for another example where ferromagnetic materials are used to probe an axion wind, although at much higher masses than the range considered in this work.)

If axionlike particles make up the DM, then they will also be non-relativistic.  This implies that $a$ oscillates at its Compton frequency $f_a=m_ac^2/2\pi\hbar$, namely
\begin{equation}\label{eq:axwave}
    a(\bm x,t)\approx a_0(\bm x)\cos(2\pi f_at).
\end{equation}
Moreover, the spatial gradients of $a$ are suppressed by its velocity $v_\DM\sim10^{-3}c$, so that
\begin{equation}
    \nabla a\sim\frac{\bm v_\DM}{c^2}\sqrt{2\rho_\DM}\sin(2\pi f_at),
\end{equation}
where $\rho_\DM\approx0.3\,\mathrm{GeV/cm}^3$ is the local DM energy density~\cite{Evans:2018bqy}.  The effective magnetic field in \eqref{Baxion} is then an AC field with frequency $f_a$ and amplitude
\begin{equation}\label{eq:axelectron_amplitude}
    B_{ae}\sim g_{ae}\cdot4\times10^{-8}\,\mathrm T.
\end{equation}
Note that the monochromatic time dependence in \eqref{axwave} only applies on timescales shorter than the coherence time $t_\mathrm{coh}\sim c^2/f_av_\DM^2$.  On longer timescales, the amplitude $a_0$ and gradient of the axion will vary stochastically~\cite{Foster_2018,Centers:2019dyn,Lisanti_2021,Blinov:2024jiz,Cheong:2024ose}.%
\footnote{If the integration time $t_\mathrm{int}$ of the experiment exceeds one day, then the direction of the gradient will also precess (in the frame of the experiment) due to the rotation of the Earth.  Such effects must be accounted for in the data analysis (see e.g. \citeR{Fedderke_2021,Fedderke_2021analysis}), but will not affect the overall sensitivity of the setup.  A similar effect will occur for the DPDM direction in the case of \secref{dark-photon}.}
Equivalently, in frequency space, the AC signal will be peaked at $f_a$, but exhibit a linewidth $\sim10^{-6}f_a$.

In the top left plot of \figref{projections}, we show the projected sensitivity of the setups described in \tabref{parameters} to an axion-electron coupling for $t_\mathrm{int}=1\,\mathrm{yr}$ of integration time.  As shown in \appref{SNR}, the signal-to-noise ratio (SNR) for a given setup is
\begin{equation}\label{eq:SNR}
    \mathrm{SNR}=\frac{B_{ae}^2}6\sqrt{\mathrm{Tr}\left[\left(S_{BB}^\mathrm{tot}\right)^{-2}\right]\cdot t_\mathrm{int}\cdot\min(t_\mathrm{int},t_\mathrm{coh})},
\end{equation}
where $B_{ae}$ is the amplitude in \eqref{axelectron_amplitude}, and the last factor accounts for the incoherence of the signal when $t_\mathrm{int}>t_\mathrm{coh}$.  In all our projections, we set $\mathrm{SNR}=3$.  In \figref{projections}, we show a number of existing constraints on $g_{ae}$, including the following: limits based on old comagnetometer data~\cite{Lee2023}, constraints on axion-mediated forces from a torsion pendulum experiment~\cite{terrano2015}, limits on solar axions from XENONnT electronic recoil data~\cite{Aprile_2022}, and constraints based on the brightness of the tip of the red-giant branch~\cite{Capozzi_2020}.  Laboratory-based constraints (comagnetometers, torsion pendulum, and XENONnT) are shown in darker shades of gray, while astrophysical ones (tip of the red-giant branch) are shown in lighter shades.  \figref{projections} shows that even an existing levitated ferromagnet setup can be competitive with the limits from comagnetometer or torsion pendulum experiments, while a future levitated or freefall setup can surpass all existing probes of an axion-electron coupling for $m_a\lesssim10^{-15}\,\mathrm{eV}$.

\subsection{Dark-photon kinetic mixing}
\label{sec:dark-photon}

A kinetically mixed dark photon $A'_\mu$, with mass $m_{A'}$, is a vector boson which may mix with the SM photon.  There are multiple equivalent descriptions of the interaction between the dark photon and SM photon (see Appendix~A of \citeR{Fedderke_2021} for further discussion), but the most useful for very low dark-photon masses is via the operator
\begin{equation}
    \LL_{A'}\supset\frac\varepsilon{\mu_0}\left(\frac{m_{A'}c}\hbar\right)^2A_\mu A'^\mu.
\end{equation}
We see that if $A'_\mu$ is treated as a background field, then it has an effect equivalent to a current
\begin{equation}
    J_\mathrm{eff}^\mu=-\frac\varepsilon{\mu_0}\left(\frac{m_{A'}c}\hbar\right)^2A'^\mu.
\end{equation}

Much like the case of axion DM, if dark photons make up the DM, they will be non-relativistic, and so should have negligible spatial gradients and oscillate at their Compton frequency $f_{A'}$.  Moreover, because its equations of motion necessitate $\partial_\mu A'^\mu=0$, then the DPDM should have $A'^0=0$ (i.e., no effective charge).  In this case,
\begin{equation}\label{eq:DPDM_wave}
    \bm A'(\bm x,t)\approx\mathrm{Re}\left[\sum_{i=x,y,z}A'_{i,0}e^{-2\pi if_{A'}t}\right],
\end{equation}
where $A'_{i,0}$ are complex amplitudes for each spatial component of $\bm A'_0$ (which may have independent phases, so that $\bm A'$ can be elliptically polarized; see footnote~\ref{ftnt:linear_pol}.).

The (spatial components of the) effective current $\bm J_\mathrm{eff}$ will also be approximately constant throughout space and oscillate at frequency $f_{A'}$.  This effective current will generate observable electromagnetic fields through the Amp\`ere-Maxwell law
\begin{equation}\label{eq:electric}
    \nabla\times\bm B-\frac{\partial_t\bm E}{c^2}=\mu_0\bm J_\mathrm{eff}.
\end{equation}
The electric field term in \eqref{electric} can be ignored in contexts where the Compton wavelength $\lambda_{A'}=c/f_{A'}$ of the dark photon is much larger than the size of the experimental apparatus $L$.%
\footnote{More specifically, by ``apparatus" here, we mean the size of the conducting shield which sets the electric field boundary conditions; see \citeR[s]{maglev,Fedderke_2021,Bloch_2024} for further discussion.}
This implies that the primary observable effect of the dark photon is an oscillating magnetic field.  Generically, this magnetic field will have amplitude
\begin{align}\label{eq:DPDM_signal}
    B_{A'}&\sim \mu_0J_\mathrm{eff}L\sim\frac{\sqrt{2\mu_0\rho_\DM}c}\hbar\varepsilon m_{A'}L\\
    &\sim7\times10^{-21}\,\mathrm{T}\left(\frac\varepsilon{10^{-8}}\right)\left(\frac{f_{A'}}{30\,\mathrm{Hz}}\right)\left(\frac L{10\,\mathrm{cm}}\right).
\end{align}
Note that by symmetry, $\bm B_{A'}$ generically vanishes at the center of the apparatus~\cite{maglev}, and so the ferromagnet should be located off-center within the apparatus in order to experience a nonzero DPDM-induced magnetic field.  In scenarios where the ferromagnet is levitated above a superconductor, this will typically be satisfied, as the ferromagnet will be much closer to the floor than the ceiling of the apparatus.

In the top right plot of \figref{projections}, we show the projected sensitivity of ferromagnets to DPDM.  The existing DPDM constraints shown include limits from: global unshielded magnetometer data maintained by the SuperMAG collaboration~\cite{Fedderke_2021,Fedderke_2021analysis,friel2024}; unshielded magnetometer measurements made by the SNIPE Hunt collaboration~\cite{sulai2023hunt}; magnetometer measurements taken inside a shielded room by the AMAILS collaboration~\cite{jiang2023search}; non-observation of CMB-photon conversion into (non-DM) dark photons by the FIRAS instrument~\cite{Caputo:2020bdy}; heating of the dwarf galaxy Leo T~\cite{Wadekar:2019xnf}; and resonant conversion of DPDM during the dark ages~\cite{McDermott:2019lch}.  A future levitated setup could become the leading probe of DPDM across the entire mass range shown in \figref{projections}.  Additionally, a freefall setup could be competitive with even the leading astrophysical constraint (Leo T) at low masses $m_{A'}\lesssim10^{-16}\,\mathrm{eV}$.

\subsection{Axion-photon coupling}
\label{sec:axion-photon}

In addition to the coupling to electrons described by \eqref{axion_electron}, an axionlike particle may also exhibit a coupling to photons via the operator
\begin{equation}\label{eq:axion_photon}
    \LL\supset\frac{g_{a\gamma}\sqrt{\hbar c^3}}{4\mu_0}aF_{\mu\nu}\widetilde F^{\mu\nu},
\end{equation}
where $\widetilde F^{\mu\nu}=\frac12\epsilon^{\mu\nu\rho\sigma}F_{\rho\sigma}$.  Similar to the DPDM case, this operator is equivalent to an effective current
\begin{equation}
    J_\mathrm{eff}^\mu=-\frac{g_{a\gamma}\sqrt{\hbar c^3}}{\mu_0}\partial_\nu a\widetilde F^{\mu\nu}.
\end{equation}
Again taking the axion DM ansatz in \eqref{axwave} [with negligible spatial gradients], we find $J_\mathrm{eff}^0=0$ and spatial components
\begin{equation}\label{eq:axion_Jeff}
    \bm J_\mathrm{eff}=\sqrt{\frac{c^5}\hbar}\frac{g_{a\gamma}m_aa_0}{\mu_0}\bm B_0\sin(2\pi f_at)
\end{equation}
One crucial difference from the DPDM case is that the effective current in \eqref{axion_Jeff} requires the presence of a background magnetic field $\bm B_0$.  In our case, the magnetic field of the ferromagnet itself can act as $\bm B_0$!  (See also \citeR{Gramolin_2021} for another example where the magnetic field from a ferromagnet is used to induce axion-photon conversion.)

Unlike the DPDM case, the current in \eqref{axion_Jeff} will not be uniform, and so computing the resulting AC magnetic field $\bm B_{a\gamma}$ is more complicated.  Generically, this must be evaluated numerically, but in \appref{axion}, we parametrically estimate it as
\begin{align}\label{eq:axion_estimate}
    &B_{a\gamma}\sim\mathcal O(0.1)\cdot\sqrt{2\hbar c\rho_\DM}\mu_0\frac{g_{a\gamma}\mu}{L^2}\\
    &\sim3\times10^{-21}\,\mathrm{T}\left(\frac{g_{a\gamma}}{10^{-9}\,\mathrm{GeV}^{-1}}\right)\left(\frac\mu{20\,\mathrm{mA\cdot m}^2}\right)\left(\frac{10\,\mathrm{cm}}L\right)^2,
\end{align}
where $\mu$ is the magnetic moment of the ferromagnet, and $L$ is again the size of the experimental apparatus.

\begin{figure*}[t]
\includegraphics[width=0.49\textwidth]{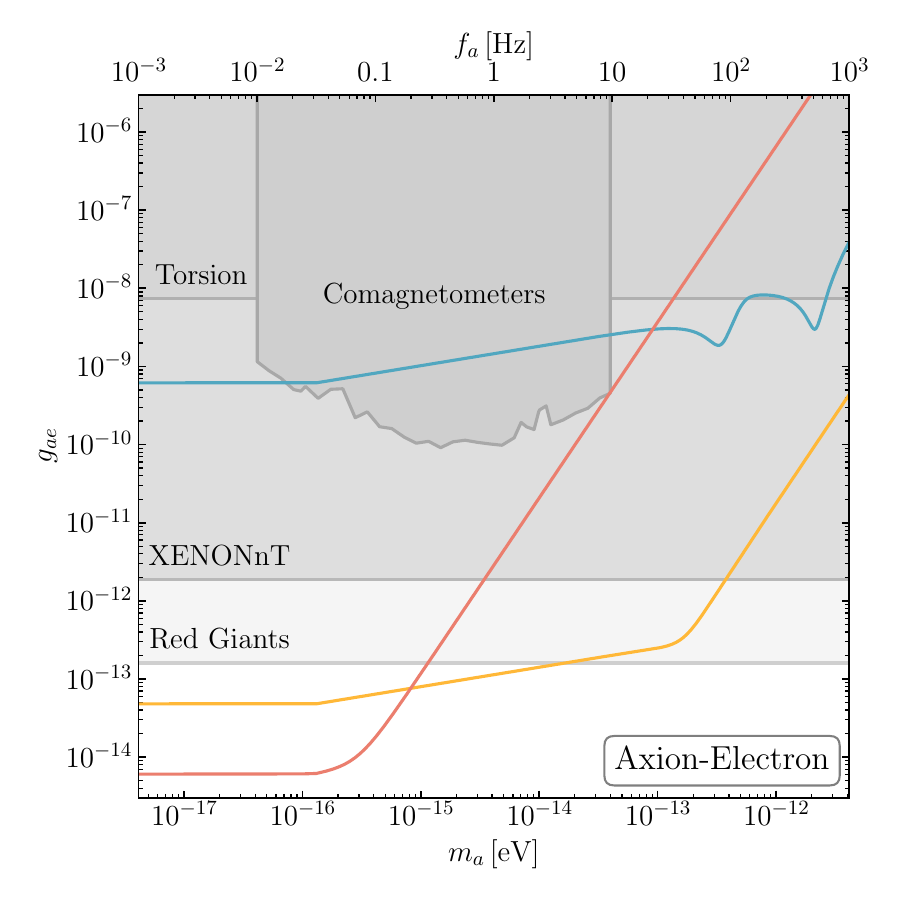}
\includegraphics[width=0.49\textwidth]{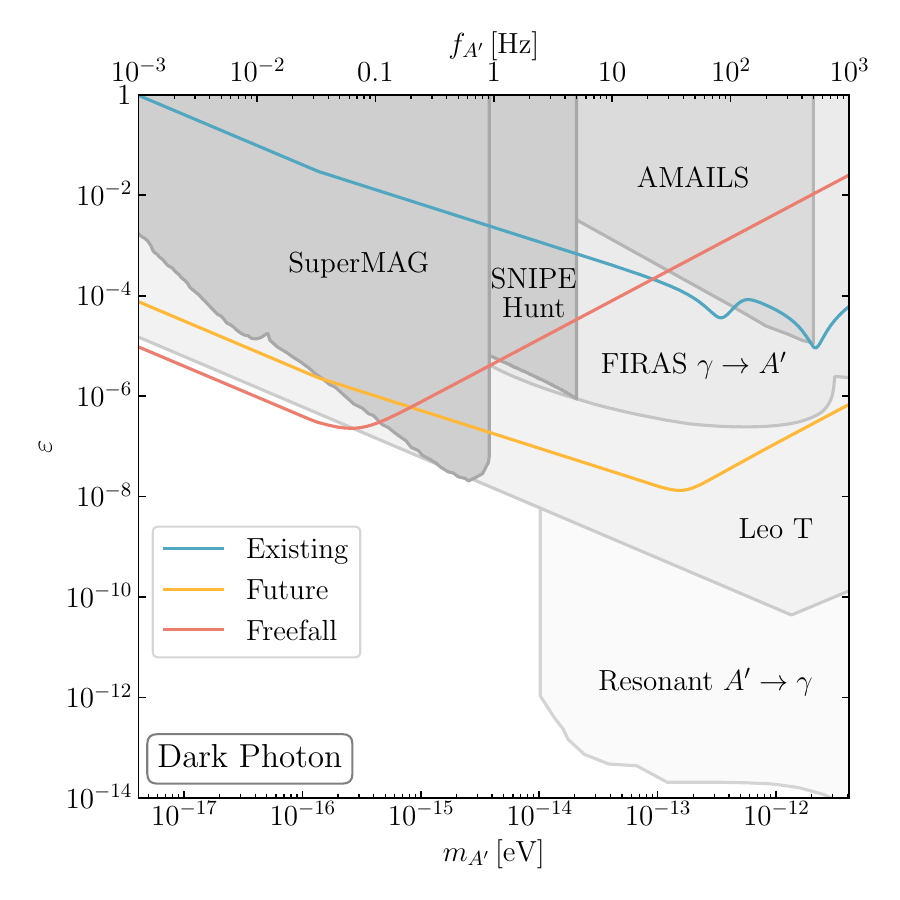}
\includegraphics[width=0.49\textwidth]{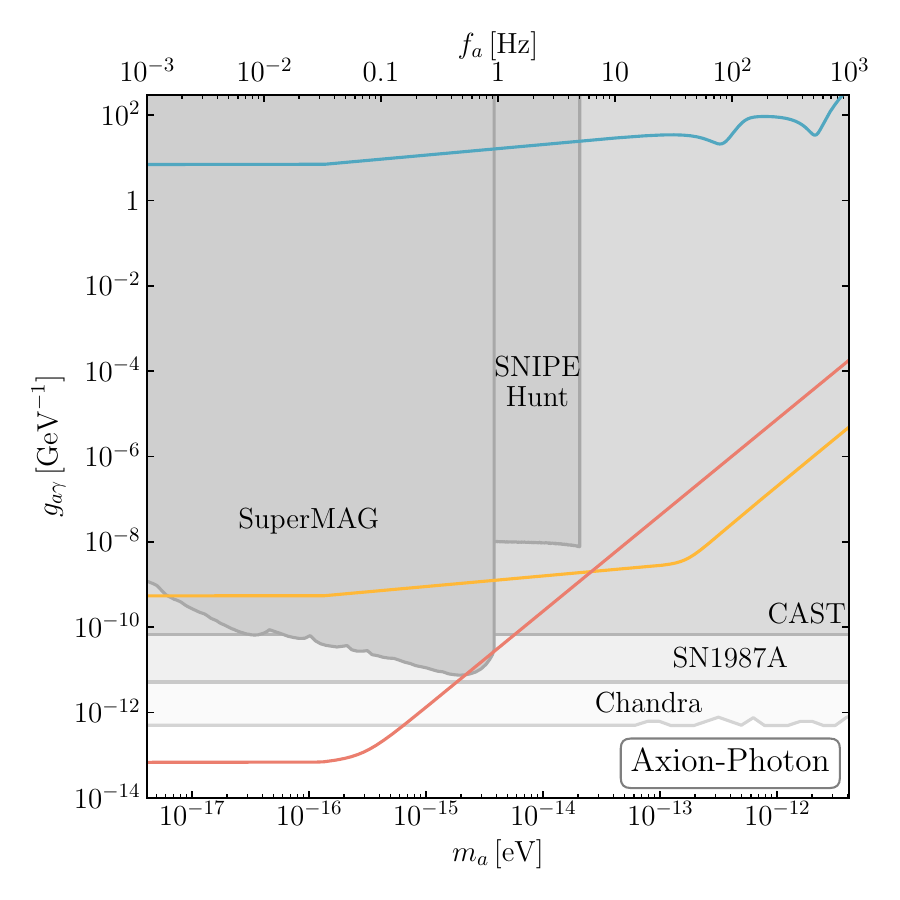}
\caption{\label{fig:projections}%
    Projected sensitivities of ferromagnets to an axion-electron coupling $g_{ae}$, DPDM kinetic mixing $\varepsilon$, and an axion-photon coupling $g_{a\gamma}$.  In each case, we show three projections corresponding to the parameter choices in \tabref{parameters}.  In all cases, we take an integration time of $t_\mathrm{int}=1\,\mathrm{yr}$ and set $\mathrm{SNR}=3$.  In the DPDM and axion-photon cases, we take an apparatus size of $L=10\,\mathrm{cm}$.  In darker shades of gray, we show existing laboratory constraints for these various models, while in lighter shades we show astrophysical constraints (see respective section of \secref{DM} for descriptions).%
    \footnote[2]{Several of these limits were acquired from \citeR[s]{Oharegithub,Caputo:2021eaa}.  See also \citeR[s]{Cardoso_2018,Witte:2020rvb,Caputo:2020bdy,escudero2023axion} for other limits in this mass range which are not shown here, and \citeR{Bloch_2024} for a brief discussion of the caveats regarding those limits.}
    We see that in all cases, ferromagnets can be the strongest laboratory probe of ultralight DM across a broad range of masses, and in some cases can even surpass astrophysical constraints.}
\end{figure*}

In the bottom plot in \figref{projections}, we show the projected sensitivity of ferromagnets to an axion-electron coupling.  The existing constraints on $g_{a\gamma}$ shown include limits from SuperMAG~\cite{Arza_2022,friel2024}, SNIPE Hunt, the CAST helioscope search for solar axions~\cite{Anastassopoulos:2017ftl}, non-observation of gamma rays in coincidence with SN1987A~\cite{Hoof_2023}, and X-ray observations of the quasar H1821+643 from the Chandra telescope~\cite{sisk-reynes2021}.  Note that in the case of an axion-photon coupling, a freefall setup may have significantly better sensitivity than a levitated setup.  This is because the background magnetic field $\bm B_0$ is sourced by the ferromagnet itself.  A ferromagnet in freefall can be much larger (and so source $\bm B_0$ over a larger volume) than a ferromagnet levitated over a superconductor because, in the latter case, the size is constrained by the critical field of the superconductor [see \eqref{surface_field}].  In fact, a freefall setup can even be more sensitive than all existing constraints at low masses $m_a\lesssim10^{-15}\,\mathrm{eV}$.

We note that in levitated setups, it may be possible to apply an additional magnetic field to act as $\bm B_0$ in order to enhance the sensitivity to an axion-photon coupling [although this will affect the trapping potential $V(\bm x,\N)$].  We leave further exploration of this idea to future work.

\section{Conclusion}
\label{sec:conclusion}

In this work, we determined the sensitivity of levitated ferromagnets to AC magnetic fields and to various ultralight DM candidates.  In the presence of an applied magnetic field, a ferromagnet may either precess around the applied field (similar to an electron spin) or librate in the plane of the applied field (similar to a compass needle).  While the distinction between these behaviors has been studied for DC magnetic fields, the cases when precession v.s. libration occurs in the presence of an AC magnetic field has not been adequately studied.  In this work, we determined the response of a ferromagnet to an AC magnetic field as a function of frequency, paying special attention to the presence of any trapping potential used to levitate the ferromagnet.  We determined three possible cases for the behavior of the system: a ``trapped" case, where the trapping potential is strong in both directions so that only libration occurs; a ``partially trapped" case, where the ferromagnet is only trapped strongly in one direction, so precession is possible in one direction within a certain frequency range; and a ``gyroscope" case, where the ferromagnet is weakly trapped, so that precession can occur in both directions within a certain frequency range.

We then computed the magnetic-field sensitivity of various ferromagnet setups, using the formalism of \secref{ferromagnet} to account for motion in both angular modes.  We considered three possible setups: one representative of an existing levitated setup~\cite{ahrens2024} (but with an additional readout for the $\theta$-mode), a future levitated setup, and a space-based freefall setup comparable to LISA Pathfinder.  All three setups manifest the ``trapped" behavior.  In \eqref{therm_dom_eta}, we show the optimal range for the readout coupling.  This range comes from demanding that thermal noise dominates over imprecision and back-action noise at low frequencies.  (This is only possible when \eqref{therm_dom} is met; see \secref{sources} for optimal choices when this condition is not met.)  Both modes of the existing setup do not fall in the range in \eqref{therm_dom_eta}, and so are under-coupled.  The other two lie in the range in \eqref{therm_dom_eta}, but the freefall setup could benefit from an even stronger coupling, which would improve its sensitivity at high frequencies.  We also note that the freefall setup could be further improved if the system can be lowered to cryogenic temperatures and ultrahigh vacuum, similar to Gravity Probe B~\cite{GravityProbeB,gps2021}.

Finally, we use the results of \secref{sensitivity} to determine the sensitivity of these setups to various DM candidates.  We consider sensitivity to an axion-electron coupling, a dark-photon kinetic mixing, and an axion-photon coupling.  While many experiments which detect magnetic fields have sensitivity to either an axion-electron coupling~\cite{Lee2023} or to a kinetic mixing and an axion-photon coupling~\cite{Fedderke_2021,Arza_2022,sulai2023hunt,maglev}, levitated ferromagnets are unique in their ability to achieve good sensitivity to all three of these potential DM couplings.  In all three cases, ferromagnet setups could become the most sensitive laboratory probes of these DM candidates, and for the axion DM cases, they could surpass even the leading astrophysical constraints at low frequencies.  Levitated ferromagnets may also be sensitive to gravitational waves~\cite{Carney:2024zzk}.  We leave further exploration of this idea to future work.  While the results of this work are already quite promising, further optimization of the setups proposed here may lead to even better detection prospects for new physics.

\acknowledgments

We thank Yifan Chen, Raymond Co, Daniel Gavilan-Martin, Keisuke Harigaya, Arne Wickenbrock, and Yue Zhao for their helpful discussions.

S.K. and Z.L. are supported in part by the U.S. Department of Energy, Office of Science, National Quantum Information Science Research Centers, Superconducting Quantum Materials and Systems Center (SQMS) under Contract No. DE-AC02-07CH11359, and by the DOE Grant No. DE-SC0011842 at the University of Minnesota.  S.K. and Z.L. also acknowledge support from the Simons Foundation Targeted Grant 920184 to the Fine Theoretical Physics Institute, which positively impacted our work.  Z. L. is also supported in part by a Sloan Research Fellowship from the Alfred P.~Sloan Foundation.

D.B, D.F.J.K., W.J., A.S, C.T., H.U., and A.V. acknowledge support from the QuantERA II Programme (project LEMAQUME) that has received funding from the European Union’s Horizon 2020 research and innovation programme under Grant Agreement No 101017733.

The work of D.F.J.K. was supported by the U.S. National Science Foundation under grant PHYS-2110388.

The work of A.O.S. was supported by the U.S. National Science Foundation CAREER grant PHY-2145162, and by the Gordon and Betty Moore Foundation, grant DOI 10.37807/gbmf12248.

C.T. and H.U. would like to thank for support the UKRI EPSRC (EP/W007444/1, EP/V035975/1, EP/V000624/1 and EP/X009491/1), the Leverhulme Trust (RPG-2022-57), and the EU Horizon Europe EIC Pathfinder project QuCoM (10032223).

A.V. acknowledges financial support from the Italian Ministry for University and Research within the Italy-Singapore Scientific and Technological Cooperation Agreement 2023-2025.

T.W. acknowledges support from  A*STAR Career Development Fund (222D800028), Italy-Singapore science and technology collaboration grant (R23I0IR042), Delta-Q (C230917004, Quantum Sensing), and Competitive Research Programme (NRF-CRP30-2023-0002).

The code used for this research is made publicly available through Github~\cite{github} under CC-BY-NC-SA.

\bibliographystyle{JHEP}
\bibliography{references.bib}

\clearpage
\appendix

\renewcommand{\theequation}{A-\arabic{equation}}
\renewcommand{\thefigure}{A-\arabic{figure}}
\setcounter{equation}{0}
\setcounter{figure}{0}
\section{Signal-to-noise ratio}
\label{app:SNR}

In this appendix, we derive an appropriate SNR for our system.  Defining such an SNR is complicated by the fact that we have sensitivity to magnetic fields in two directions, so in this appendix, we pay special attention to the matrix nature of our noise $S_{BB}^\mathrm{tot}(\omega)$.  Here, we consider sensitivity to an AC magnetic field signal%
\footnote{This would be the form of the signal in the case of an axion-electron coupling or DPDM kinetic mixing.  In the case of an axion-photon coupling, each term in \eqref{ACsignal} would have the same phase $\Phi_S$, so that the signal is linearly polarized [see footnote \ref{ftnt:linear_pol}].  Moreover, the direction of the signal is, in principle, not random but rather can be predicted through a sufficiently accurate signal calculation.  As we do not perform such a calculation in this work, we treat the direction as random in our sensitivity projections, and so we still apply the formalism of this appendix.}
\begin{align}\label{eq:ACsignal}
    \bm B_S(t)&=B_{S,n}\cos(\omega_St+\Phi_{S,n})\N_0\nl
    +B_{S,\theta}\cos(\omega_St+\Phi_{S,\theta})\THETA_0\nl
    +B_{S,\phi}\cos(\omega_St+\Phi_{S,\phi})\PHI_0
\end{align}
of unknown direction and phase.  (As we are insensitive to the $\N_0$ direction, we will ignore $B_{S,n}$.)  We will take the distribution of $\bm B_S$ to be Gaussian and isotropic so that each component follows an independent Gaussian distribution with mean zero and $\langle B_{S,\alpha}^2\rangle=\overline B_S^2/3$ (where $\langle\cdot\rangle$ represents an ensemble average).  Meanwhile, the noise $\bm B_N(t)$ has a Fourier transform whose components satisfy
\begin{equation}
    \langle\tilde B_{N,\alpha}(\omega)\tilde B_{N,\beta}(\omega)^*\rangle=\frac{S^\mathrm{tot}_{BB,\alpha\beta}(\omega)t_\mathrm{int}}2,
\end{equation}
where $t_\mathrm{int}$ is the integration time of the experiment (and $\tilde{\bm B}_N$ is uncorrelated at different frequencies).

When we perform an experiment, we measure the Fourier transform of the total magnetic field $\tilde{\bm B}_\mathrm{tot}=\tilde{\bm B}_S+\tilde{\bm B}_N$, if there exists a signal, or simply $\tilde{\bm B}_\mathrm{tot}=\tilde{\bm B}_N$, if there does not.  To distinguish these two scenarios, we ought to combine the information from the different components of $\tilde{\bm B}_\mathrm{tot}$ (for a fixed frequency) into a single test statistic
\begin{equation}
    q=\tilde{\bm B}_\mathrm{tot}^\dagger X\tilde{\bm B}_\mathrm{tot},
\end{equation}
for some Hermitian matrix $X$ to be chosen momentarily.  In the scenario where there is no signal, this statistic has
\begin{align}
    \langle q\rangle_0&=\left\langle\tilde{\bm B}_N^\dagger X\tilde{\bm B}_N\right\rangle=\mathrm{Tr}\left[X\left\langle\tilde{\bm B}_N\tilde{\bm B}_N^\dagger\right\rangle\right]\\
    &=\frac{t_\mathrm{int}}2\mathrm{Tr}\left[XS_{BB}^\mathrm{tot}\right],\\
    \langle q^2\rangle_0&=\left\langle\tilde{\bm B}_N^\dagger X\tilde{\bm B}_N\tilde{\bm B}_N^\dagger X\tilde{\bm B}_N\right\rangle\\
    &=\frac{t_\mathrm{int}^2}4\left(\mathrm{Tr}\left[XS_{BB}^\mathrm{tot}\right]^2+\mathrm{Tr}\left[XS_{BB}^\mathrm{tot}XS_{BB}^\mathrm{tot}\right]\right).
\end{align}
On the other hand, when a signal is present, its expectation is
\begin{align}
    \langle q\rangle_S&=\left\langle\tilde{\bm B}_S^\dagger X\tilde{\bm B}_S\right\rangle+\left\langle\tilde{\bm B}_N^\dagger X\tilde{\bm B}_N\right\rangle\\
    &=\frac{\overline B_S^2t_\mathrm{int}^2}{12}\mathrm{Tr}\left[X\right]+\frac{t_\mathrm{int}}2\mathrm{Tr}\left[XS_{BB}^\mathrm{tot}\right].
\end{align}
A signal is distinguishable when the difference between $q$ with/without the signal exceeds the standard deviation of $q$ without any signal.  That is, we should define the SNR as
\begin{align}
    \mathrm{SNR}&=\frac{\langle q\rangle_S-\langle q\rangle_0}{\sqrt{\langle q^2\rangle_0-\langle q\rangle_0^2}}\\
    &=\frac{\overline B_S^2t_\mathrm{int}\mathrm{Tr}\left[X\right]}{6\sqrt{\mathrm{Tr}\left[XS_{BB}^\mathrm{tot}XS_{BB}^\mathrm{tot}\right]}}.
\end{align}

Now we can consider what an optimal choice of $X$ would be.  The only matrix structure available is $S_{BB}^\mathrm{tot}$, and so we should choose $X\propto(S_{BB}^\mathrm{tot})^n$ for some $n$.  If $S_{BB}^\mathrm{tot}$ has two eigenvalues $\lambda_1,\lambda_2$, then this becomes
\begin{equation}
    \mathrm{SNR}=\frac{\overline B_S^2t_\mathrm{int}(\lambda_1^n+\lambda_2^n)}{6\sqrt{\lambda_1^{2n+2}+\lambda_2^{2n+2}}}.
\end{equation}
It is not difficult to show that this expression is maximized for $n=-2$, leading to an optimal SNR of
\begin{align}
    \mathrm{SNR}&=\frac{\overline B_S^2t_\mathrm{int}}6\sqrt{\lambda_1^{-2}+\lambda_2^{-2}}\\
    &=\frac{\overline B_S^2t_\mathrm{int}}6\sqrt{\mathrm{Tr}\left[\left(S_{BB}^\mathrm{tot}\right)^{-2}\right]}.
\end{align}
This can be equivalently phrased as computing the two SNRs representing the sensitivity along each eigenvector, and then summing them in quadrature.

Finally, we note that the above discussion applies when the signal is entirely coherent throughout the duration of the experiment.  If the coherence time of the signal is shorter than the duration of the experiment, then we can consider each coherence time as an independent experiment.  In this case, the SNRs for the individual experiments can be summed in quadrature, so that the total SNR is
\begin{equation}
    \mathrm{SNR}=\frac{\overline B_S^2t_\mathrm{coh}}6\sqrt{\mathrm{Tr}\left[\left(S_{BB}^\mathrm{tot}\right)^{-2}\right]}\cdot\sqrt{\frac{t_\mathrm{int}}{t_\mathrm{coh}}}.
\end{equation}

\renewcommand{\theequation}{B-\arabic{equation}}
\renewcommand{\thefigure}{B-\arabic{figure}}
\setcounter{equation}{0}
\setcounter{figure}{0}
\section{Axion-photon coupling signal}
\label{app:axion}

In this appendix, we estimate the magnetic-field signal $\bm B_{a\gamma}$ induced by axion DM which couples to photons.  This computation largely follows Appendix B of \citeR{maglev}, but the background magnetic field $\bm B_0$ in this case will be sourced by a magnetic dipole (the ferromagnet) instead of a pair of anti-Helmholtz coils.  The axion DM signal is given by~\cite{Chaudhuri:2014dla,maglev}
\begin{equation}\label{eq:cavity_modes}
    \bm B_{a\gamma}(\bm r)=\sum_nc_n\frac{f_n}{f_a}\bm B_n(\bm r)e^{-im_at},
\end{equation}
\begin{equation}\label{eq:overlap}
    c_n=-\frac{\sqrt{\hbar c^3}g_{a\gamma}f_a^2a_0}{f_n^2-f_a^2}\int dV\,\bm E_n(\bm r)^*\cdot\bm B_0(\bm r),
\end{equation}
where $\bm E_n$ and $\bm B_n$ are the electric/magnetic fields for the cavity modes of the shield (which have corresponding frequencies $f_n$ and are normalized so that $\int dV|\bm E_n|^2=1$).

In order to estimate the overlap integral in \eqref{overlap}, it is useful to write $\bm B_0$ as the gradient of a magnetic potential, which is possible in the absence of free currents or magnetization (for static magnetic fields).  Of course, the ferromagnet is magnetized, so $\bm B_0$ itself cannot be written this way.  However because there are no free currents, we may write $\bm B_0=\mu_0(\bm H_0+\bm M)$ with $\nabla\times\bm H_0=0$.  Because $\bm H_0$ is curl-free, it can be written as $\bm H_0=\nabla\Psi_0$.  Then the overlap integral becomes
\begin{equation}\label{eq:integrals}
    \int dV\,\bm E_n^*\cdot\bm B_0=\mu_0\left(\int d\bm A\cdot\bm E_n\Psi_0+\int dV\,\bm E_n^*\cdot\bm M\right).
\end{equation}
The second integral in \eqref{integrals} only has support over the volume of the ferromagnet.  Assuming that the ferromagnet is much smaller than the size of the shield, $\bm E_n$ will be roughly constant over this volume, and so we can approximate this integral as $\mu_0\bm E_n^*\cdot\bm\mu$, where $\bm\mu$ is the magnetic moment of the ferromagnet.

To compute the first integral in \eqref{integrals} requires an exact expression for $\Psi_0$ on the boundary of the shield.  In the absence of the shield, this is just a magnetic dipole.  However, if the shield is superconducting, then $\bm B_0$ will be modified in order to ensure that the perpendicular magnetic field vanishes at the boundary of the shield.  As in \secref{potential}, this can be accounted for via the method of images.  As we only wish to derive a parametric estimate for the axion DM signal, we will simply take the magnetic potential of a dipole
\begin{equation}
    \Psi_0(\bm r)=\frac{\bm\mu\cdot(\bm r-\bm r_0)}{4\pi|\bm r-\bm r_0|^3},
\end{equation}
where $\bm r_0$ is the location of the ferromagnet.  Parametrically, the first integral in \eqref{integrals} is then also $\sim\mu_0\bm E_n^*\cdot\bm\mu$.

Since $E_n,B_n\sim L^{-3/2}$ and $f_n\sim\frac cL\gg f_a$, then parametrically \eqref{cavity_modes} becomes
\begin{align}
    B_{a\gamma}&\sim\frac{\sqrt{\hbar c^3}g_{a\gamma}f_aa_0}{f_n}\cdot\mu_0E_n\mu\cdot B_n\\
    &\sim\sqrt{2\hbar c\rho_\DM}\mu_0\frac{g_{a\gamma}\mu}{L^2}.
\label{eq:axion_parametric}\end{align}
In principle, the overlap integrals in \eqref{integrals} can be computed for each mode, and they can be summed to determine the exact proportionality constant in \eqref{axion_parametric}.  Numerically, we find that this sum exhibits poor convergence, and so we remain content with a parametric estimate.  (In \eqref{axion_estimate}, we include a conservative $\mathcal O(0.1)$ factor, in line with the factor computed in \citeR{maglev}.)  In future work, detailed finite element method calculations may be necessary to predict an accurate axion DM signal.

\end{document}